\documentclass[
    reprint,
    superscriptaddress,
    amsmath,
    amssymb,
    aps,
    prl
]{revtex4-2}

\usepackage{graphicx}
\usepackage{dcolumn}
\usepackage{bm}
\usepackage{physics}
\usepackage{xcolor}
\usepackage{hyperref}
\hypersetup{
    colorlinks=true,
    linkcolor=blue,
    citecolor=blue,
    urlcolor=blue
}

\begin{document}

\title{Ferron polariton Hall effect:\\\ Deflecting optically rectified polarization with magnetic fields}

\author{Omer Yaniv}
\email{yaniv@mail.tau.ac.il}
\affiliation{School of Physics and Astronomy, Tel Aviv University,
Tel Aviv 6997801, Israel}

\author{Dominik M. Juraschek}
\email{dominik.juraschek@unifr.ch}
\affiliation{School of Physics and Astronomy, Tel Aviv University,
Tel Aviv 6997801, Israel}
\affiliation{Department of Applied Physics and Science Education,
Eindhoven University of Technology, 5612 AP Eindhoven, Netherlands}
\affiliation{Department of Physics, University of Fribourg, CH-1700 Fribourg, Switzerland}
\affiliation{Paul Scherrer Institut, CH-5232 Villigen PSI, Switzerland}

\date{\today}


\begin{abstract}
Ferrons are elementary excitations of ferroelectric order that can propagate and transport polarization when excited with light.
 Here, we show theoretically that ferron polaritons exhibit a Hall effect: an applied magnetic field generates polarization transport perpendicular to the propagation direction of a ferron polariton excited by an incident terahertz pulse. We demonstrate this effect for the example of ferroelectric SnTe using first-principles calculations and a microscopic theory of the phonon-polariton current. The direction and magnitude of the transverse polarization transport can be controlled by the polarization of the terahertz pulse and the applied magnetic field. Our results establish a route toward light-controlled directional transport of electric polarization in ferroelectrics.
\end{abstract}

\maketitle


\section{Introduction}

Ferrons are elementary excitations of ferroelectric order that carry a finite net polarization and can therefore mediate polarization transport.
\cite{bauer2022magnonics,tang2022excitations,bauerPolarizationTransportFerroelectrics2023,zhouSurfaceFerronExcitations2023,tangElectricAnalogMagnons2024,shenObservationFerronTransport2025,morozovskaFlexocouplinginducedPhononsFerrons2025,zhaoRoleFerronsHeat2025,Subedi2026Ferrons,pols2026lattice,tang2026multiferroic,Choe2026Ferrons,Jana2026Ferrons,zhang2026electric}.
These excitations describe the dynamics of the ferroelectric order parameter within its anharmonic potential, and can microscopically be described in terms of lattice vibrational modes, including infrared-active optical phonons carrying an oscillating electric dipole moment~\cite{pols2026lattice,tang2026multiferroic}. The net electric dipole moment of ferrons arises from a rectified lattice displacement caused by asymmetric lattice anharmonicities. Accordingly, they enable polarization transport in ferroelectrics in a similar way to how magnons enable magnetization transport in magnetic materials \cite{bauer2022magnonics,chumak2015magnon,pirro2021advances,Flebus2024MagnonicsRoadmap}.

Dipole-carrying optical phonons can further hybridize with electromagnetic radiation to form phonon polaritons, combining lattice motion with the propagation of light. When the lattice excitation carries a net electric polarization of a ferron, the resulting ferron-polariton couples polarization transport to a propagating terahertz electromagnetic field. Ferron-polariton transport has recently been predicted
and experimentally observed, establishing that polarization transport
can be launched by terahertz radiation~\cite{tang2022excitations,Choe2026Ferrons,tangElectricAnalogMagnons2024,Nursagatov2026FerronPolaritons}.

At the same time, it has been shown that lattice vibrations with elliptical polarization components, also known as chiral or axial phonons, couple to magnetic fields through their phonon magnetic moment~\cite{Juraschek2025,juraschek2:2017,Juraschek2019,Cheng2020,Geilhufe2021,Xiao2021,Ren2021,Baydin2022,Hernandez2023,Geilhufe2023,Zhang2023_BLG,Shabala2024,Klebl2024,Basini2024,Mustafa2025,Chen2025_gaugetheory,Paiva2025}. This coupling gives rise to the phonon Zeeman effect:
an applied magnetic field lifts the degeneracy between
modes of opposite handedness and therefore angular momentum. This magnetic coupling can produce transverse phonon transport, as exemplified by the phonon Hall effect, in which a
longitudinal temperature gradient generates a transverse heat
current under a magnetic field
\cite{strohm:2005,Inyushkin2007,Kagan2008,Wang2009,zhang2010topological,Zhang2011,agarwalla2011phonon,Qin2012,mori2014origin,ideue2017giant,Grissonnanche2019,Grissonnanche2020,Li2020,yang2020universal,Guo2021phononHall,chen2022large,Mangeolle2022,Sun2022,Flebus2022,Li2023phononHall,Flebus2023,Chen2024,Jin2024_PHE,Oh2024,Chen2024_PHE}.

Here, we synthesize magnetic control of lattice excitations with the optical control of ferrons. We demonstrate theoretically a ferron polariton Hall effect, in which an applied magnetic field generates polarization transport transverse to the propagation direction of a coherently excited ferron polariton. Realizing this effect requires a material that combines ferroelectricity with a sizable phonon magnetic moment. We identify ferroelectric SnTe~\cite{plekhanov2014engineering,Hernandez2023} as a promising candidate and evaluate its response by combining first-principles calculations with a microscopic theory of phonon-polariton transport. We show that both the magnitude and direction of the transverse polarization current can be controlled by the orientation and strength of the magnetic field and by the polarization of the incident terahertz pulse.


\begin{figure*}[t]
    \centering
    \includegraphics[width=0.8\textwidth]{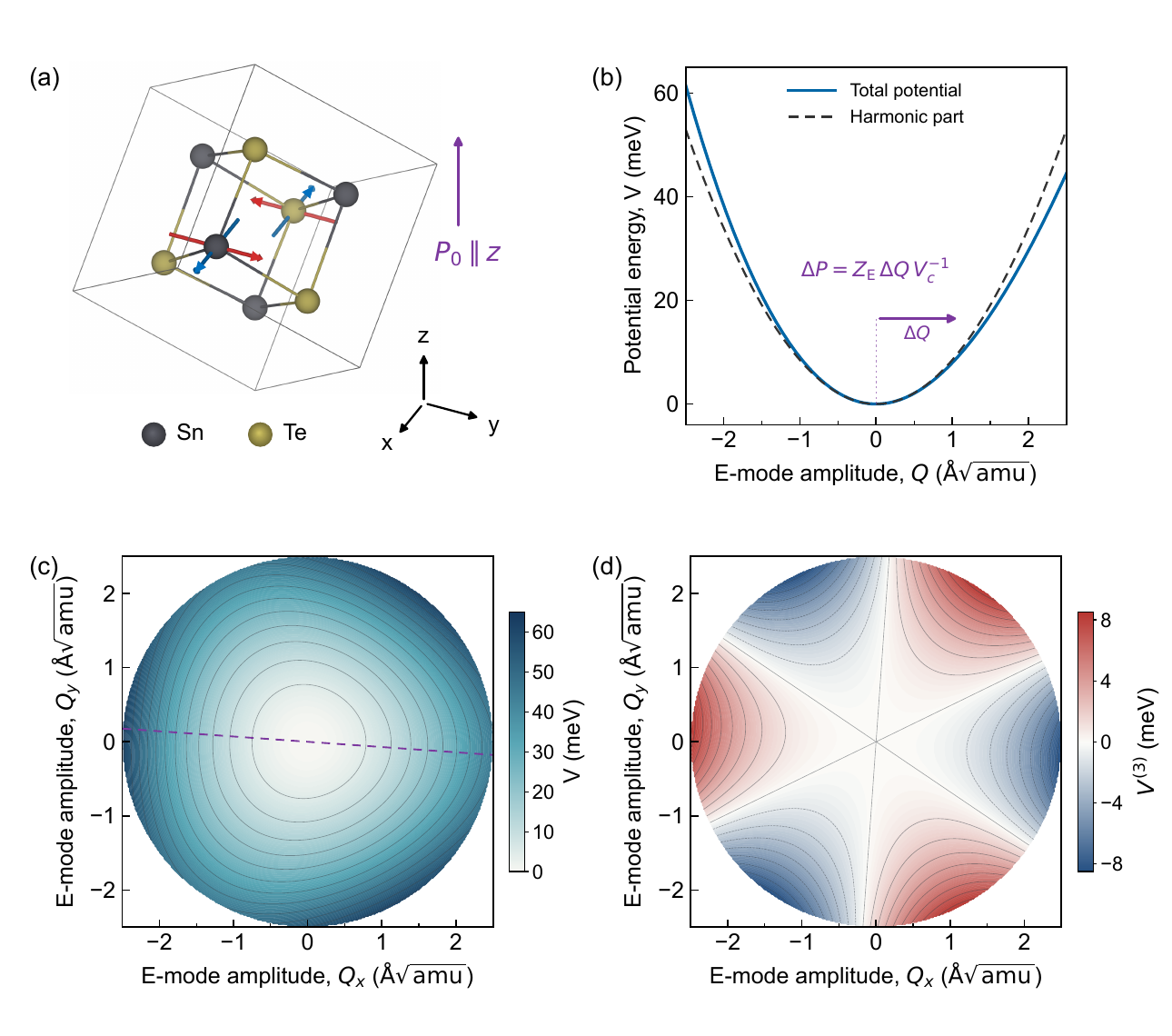}
   \caption{
(a) Ferroelectric $R3m$ structure of SnTe. The relative displacement
of the Sn and Te sublattices along the rhombohedral axis
produces the equilibrium ferroelectric polarization $\mathbf P_0$.
The degenerate $E$-mode pair describes lattice motion
perpendicular to $\mathbf P_0$, as shown in red and blue arrows.
(b) Calculated potential-energy surface along the direction
of maximal ferron polarization, with the harmonic contribution
shown as a dashed line. The cubic asymmetry rectifies
the lattice motion, producing a phonon-induced net
polarization $\Delta\mathbf P$.
(c) Full potential-energy surface in the normal coordinates
$Q_x$ and $Q_y$. The purple dashed line marks the direction of maximal
ferron polarization used for the cut in (b).
(d) Isolated cubic contribution $V^{(3)}$, showing the threefold angular
dependence imposed by the trigonal symmetry. 
}
    \label{fig:ferron}
\end{figure*}


\begin{figure*}[t]
 \centering
 \includegraphics[width=\textwidth]{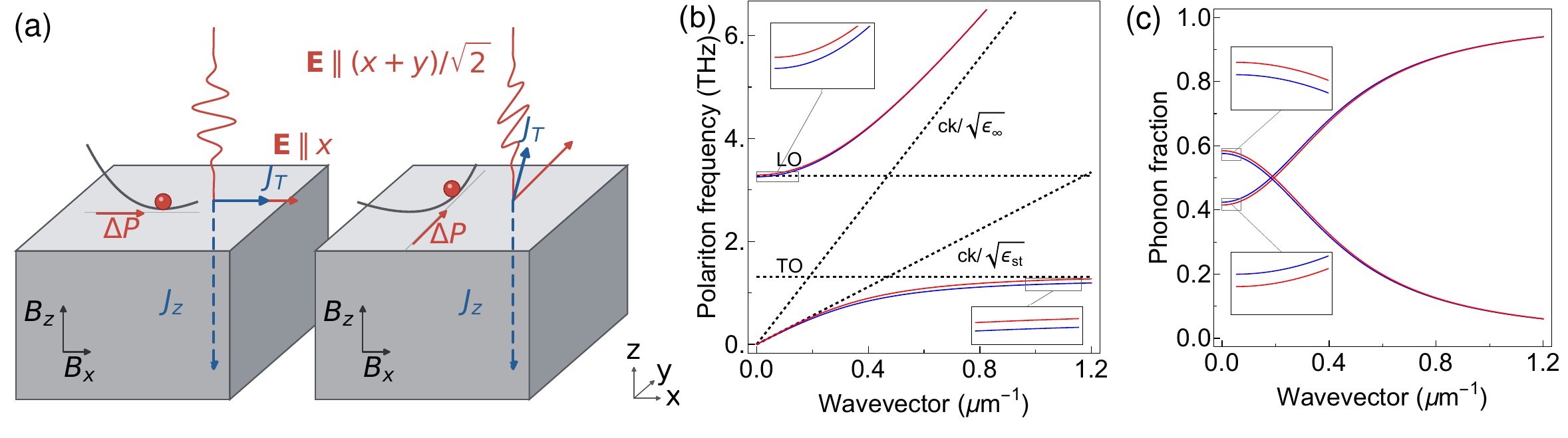}
 \caption{
 (a) A linearly polarized terahertz pulse propagating along $z$
 excites ferron polaritons carrying a net polarization
 $\Delta\mathbf P$. A magnetic field with components
 $B_x$ and $B_z$ generates a transverse polarization current
 $\mathbf J_T^P$ in addition to the longitudinal current $J_z^P$.
 The two configurations illustrate how rotating the incident pulse
 polarization redirects the net polarization and its
 transverse transport.
 (b) Dispersions of the lower and upper chiral ferron polariton
 branches, split by the phonon Zeeman effect.
 Blue and red distinguish the two opposite helicities.
 (c) Phonon fractions $F_{k,\pm}$, quantifying the lattice
 component of each polariton branch.
 The spectrum is shown for the example of $B_z=6\,\mathrm T$ and $\mu_{\mathrm{ph}}=1\,\mu_B$.
 }
 \label{fig:polariton_system}
\end{figure*}


\section{Ferron polarization in tin telluride}
SnTe is a promising material for demonstrating the ferron
polariton Hall effect by combining ferroelectric order with
infrared-active optical phonons in a material family exhibiting
giant phonon magnetic moments \cite{Baydin2022,Hernandez2023}. Large phonon Zeeman splittings,
$\hbar|\Omega_+-\Omega_-|=|\mu_{\mathrm{ph}}B|$,
have been measured in PbTe and Pb$_{1-x}$Sn$_x$Te alloys. Here, $\Omega_\pm$ denote the angular frequencies of the two oppositely circularly polarized phonon modes, and $\mu_{\mathrm{ph}}$ is the effective phonon magnetic moment governing their splitting under an applied magnetic field $B$. The measured splittings correspond to $|\mu_{\mathrm{ph}}|\simeq0.086,\mu_B$ in PbTe~\cite{Baydin2022} and values reaching $6.6,\mu_B$ in Pb$_{1-x}$Sn$_x$Te alloys~\cite{Hernandez2023}, where $\mu_B$ is the Bohr magneton.
We consider the low-temperature rhombohedral ferroelectric
phase of SnTe, with space group $R3m$ (point group $C_{3v}$),
shown in Fig.~\ref{fig:ferron}(a)
\cite{plekhanov2014engineering}.
SnTe undergoes a ferroelectric transition at
$T_c=75\,\mathrm K$ in the sample studied in
Ref.~\cite{o2017inelastic}, with the exact transition temperature
depending on carrier concentration.
The relative displacement of the Sn and Te sublattices along
the rhombohedral $[111]$ direction produces the equilibrium
ferroelectric polarization $\mathbf P_0$. We choose this
direction as the $z$ axis, with the crystal oriented accordingly
in Fig.~\ref{fig:ferron}(a).

We specifically investigate the low-frequency, doubly degenerate
optical phonon mode with $E$ symmetry, for which our first-principles
calculations yield an eigenfrequency of
$\Omega_T/(2\pi)=1.31\,\mathrm{THz}$.
The $E$ doublet describes lattice motion with electric dipole
moments perpendicular to the ferroelectric axis, which we represent
by the normal coordinates $Q_x$ and $Q_y$.
Our first-principles calculations were performed using the density functional theory framework as implemented in VASP~\cite{Kresse1996PRB,Kresse1996CMS}, within the Perdew-Burke-Ernzerhof (PBE) approximation~\cite{Perdew1996PBE}.
We adopt the two-atom rhombohedral primitive cell of
Ref.~\cite{plekhanov2014engineering}, with lattice-vector
length $a_r=4.57\,\text{\AA}$ and rhombohedral angle
$\alpha_r=58.8^\circ$, corresponding to a primitive cell
volume of $V_c=66.01\,\text{\AA}^3$.
We keep these lattice parameters fixed and relax the internal
atomic coordinates until the forces are below
$10^{-4}\,\mathrm{eV/\text{\AA}}$.
We use a plane-wave energy cutoff of $550\,\mathrm{eV}$,
an electronic convergence threshold of $10^{-8}\,\mathrm{eV}$
and a $12\times12\times12$ $k$-point mesh.
Harmonic phonons are calculated using the finite-displacement
method in Phonopy~\cite{togo2023first}, with force constants obtained from VASP
calculations in a $3\times3\times3$ supercell.
Projecting the calculated Born effective charge tensors onto
the mass-normalized $E$-mode eigenvectors yields a mode
effective charge of $Z_E=0.79\,e/\sqrt{\mathrm{amu}}$.

We next determine the net electric polarization induced along
the displacement direction of the excited $E$ mode.
The net polarization arises from  the asymmetric anharmonicity of
the potential, which rectifies the lattice motion.
To quantify this anharmonicity, we calculate the total energies
of structures displaced along combinations of the two
$E$-mode eigenvectors. Writing $Q_x=Q\cos\theta$ and
$Q_y=Q\sin\theta$, we fit these energies to the leading
anharmonic form~\cite{pols2026lattice}
\begin{equation}
    V(Q,\theta)
    =
    \frac{\Omega_T^2}{2}Q^2
    +
    V^{(3)}(Q,\theta),
    \label{eq:ferron_potential}
\end{equation}
where
$V^{(3)}(Q,\theta)=a'\cos(3\theta+\delta)Q^3$
is the cubic anharmonic contribution to the potential, whose
threefold angular dependence follows from the $C_{3v}$ symmetry. The cubic anharmonicity rectifies the lattice motion and produces
a finite time-averaged polarization~\cite{pols2026lattice,tang2026multiferroic}.
For one phonon per primitive cell, the
excitation-induced mean-square displacement is
$\langle Q^2\rangle=\hbar/\Omega_T$, yielding
\begin{equation}
    P(\theta)
    =
    -\frac{3\hbar Z_E}
           {V_c\Omega_T^3}
    a'\cos(3\theta+\delta),
    \label{eq:ferron_polarization}
\end{equation}
where $P(\theta)$ denotes the phonon-induced net
electric polarization projected along the displacement
direction of the excited $E$ mode. Its angular dependence
has the same form as that obtained for other materials
with $C_{3v}$ point group symmetry~\cite{pols2026lattice}.

For SnTe, the fit gives
$a'=0.54\,\mathrm{meV}/(\text{\AA}\sqrt{\mathrm{amu}})^3$
and $\delta=-2.9\,\mathrm{rad}$, yielding a maximum induced
net polarization of $|\Delta P|_{\max}\simeq
3.4\,\mu\mathrm C\,\mathrm{cm}^{-2}$ for one 
ferron per primitive cell, corresponding to a net dipole moment
of approximately $0.14\,e\,\text{\AA}$ per ferron. The calculated ferron polarization in SnTe exceeds that reported for the $E$ modes of LiNbO$_3$~\cite{pols2026lattice} and corresponds to a dipole per phonon comparable to that predicted for transverse optical modes in BaTiO$_3$~\cite{tang2026multiferroic}.
Figure~\ref{fig:ferron}(b) shows the calculated potential
along the direction of maximal anharmonicity, with the harmonic
contribution indicated by a dashed line. The full potential energy
surface and its cubic contribution are shown in
Fig.~\ref{fig:ferron}(c) and (d).
The net polarization $P(\theta)$ associated with the excitation identifies the $E$ mode as a ferron. The finite mode effective charge, $Z_E \neq 0$, makes the
$E$ mode infrared active, allowing it to hybridize with
terahertz radiation. The resulting phonon-polariton branches
inherit the phonon-induced net polarization through the rectified lattice component, forming propagating ferron polaritons.


\section{Ferron polariton transport}

To study the propagation and Hall effect of ferron polaritons in SnTe, we consider a linearly polarized terahertz pulse propagating along $z$ through a sample subjected to a magnetic field $\mathbf B=B_x\hat{\mathbf x}+B_z\hat{\mathbf z}$. The electric field of the pulse is
$\mathbf E(t)=E(t)(\cos\theta\,\hat{\mathbf x}
+\sin\theta\,\hat{\mathbf y})$.
The pulse coherently excites ferron polaritons
and induces a net polarization component, $\Delta\mathbf P=P(\theta)
(\cos\theta\,\hat{\mathbf x}+\sin\theta\,\hat{\mathbf y})$.
Figure~\ref{fig:polariton_system}(a) illustrates the setup
for an incident polarization along $x$ (left) and along $(x+y)/\sqrt{2}$ (right).
The two configurations demonstrate how rotating the polarization of the incident
pulse changes the induced net polarization and redirects
its transverse transport.

We next determine the polariton spectrum.
In the absence of a magnetic field, the frequencies of the
lower ($-$) and upper ($+$) ferron polariton  branches are
given by~\cite{Horton1974}
\begin{equation}
\omega_{k,\pm}^2
=
\frac{1}{2}
\left[
\Omega_L^2+\omega_\nu^2(k)
\pm
\sqrt{
\left(\Omega_L^2+\omega_\nu^2(k)\right)^2
-
4\Omega_T^2\omega_\nu^2(k)
}
\right],
\label{eq:classical_dispersion}
\end{equation}
where $\Omega_T$ and $\Omega_L$ are the transverse- and
longitudinal-optical phonon frequencies, respectively.
The bare photon dispersion is
$\omega_\nu(k)=ck/\sqrt{\varepsilon_\infty}$,
with $c$ the speed of light in vacuum and
$\varepsilon_\infty$ the high-frequency dielectric constant.
Using the calculated mode effective charge and dielectric response,
we obtain $\Omega_L/(2\pi)=3.27\,\mathrm{THz}$.

The propagating $E$ mode can be expressed as left- and right-handed
chiral modes,
$|L/R\rangle=(|x\rangle\pm i|y\rangle)/\sqrt{2}$,
carrying opposite phonon magnetic moments.
Their coupling to $B_z$ lifts the degeneracy between the two
helicities, producing chiral ferron polariton branches.
Chiral phonon polaritons have previously been investigated experimentally
in LiNbO$_3$~\cite{biggs2026ultrafast}.
The calculated dispersion in
Fig.~\ref{fig:polariton_system}(b) shows the magnetic splitting
of the lower and upper branches. We quantify the lattice component of each hybrid branch through
its phonon fraction $F_{k,\pm}$~\cite{yaniv2025phonon},
\begin{equation}
F_{k,\pm}
=
\frac{
(\Omega_L^2-\Omega_T^2)(\omega_{k,\pm}^2+\Omega_T^2)
}{
2\left[
(\omega_{k,\pm}^2-\Omega_T^2)^2
+
\Omega_T^2(\Omega_L^2-\Omega_T^2)
\right]
}
\label{eq:phonon_fraction}
\end{equation}
which we plot in Fig.~\ref{fig:polariton_system}(c).
At small wavevectors, the lower branch is predominantly photon-like and the upper branch predominantly ferron-like.
With increasing wavevector, the lower branch approaches the
transverse optical phonon frequency and becomes increasingly
ferron-like, while the upper branch becomes increasingly
photon-like. 
In addition to changing the eigenfrequencies, the magnetic field
modifies the momentum-dependent eigenvectors, giving rise to
transverse energy transport.

To calculate the phonon energy current generated by the coherent
excitation, we use the field-dependent eigenfrequencies
$\omega_{\mathbf k,\sigma}$ and normalized eigenvectors
$\boldsymbol\epsilon_{\mathbf k,\sigma}$ obtained by including the phonon Zeeman coupling in the lattice Hamiltonian (see Supplemental Material).
We then evaluate the expectation value of the energy-current
operator~\cite{sheng2006theory,yaniv2025phonon}, which contains
diagonal and off-diagonal
contributions:
\begin{equation}
\begin{aligned}
\hat J_i^E
&=
\frac{\hbar}{2V_c}
\sum_{k_z>0}
\Bigg[
\sum_{\sigma}
\omega_{\mathbf k,\sigma}
\frac{\partial\omega_{\mathbf k,\sigma}}{\partial k_i}
\hat a^\dagger_{\mathbf k,\sigma}
\hat a_{\mathbf k,\sigma}
\\
&\quad+
\frac{1}{2}\sum_{\sigma\neq\sigma'}
\sqrt{\frac{\omega_{\mathbf k,\sigma'}}
           {\omega_{\mathbf k,\sigma}}}
\left(
\omega_{\mathbf k,\sigma}^{2}
-\omega_{\mathbf k,\sigma'}^{2}
\right)
\\
&\qquad\times
\left(
\boldsymbol\epsilon_{\mathbf k,\sigma}^{*}
\cdot
\frac{\partial\boldsymbol\epsilon_{\mathbf k,\sigma'}}
     {\partial k_i}
\right)
\hat a^\dagger_{\mathbf k,\sigma}
\hat a_{\mathbf k,\sigma'}
+\mathrm{H.c.}
\Bigg].
\end{aligned}
\label{eq:Jgeneral}
\end{equation}
Here, $\hat a_{\mathbf k,\sigma}$ annihilates a ferron
in branch $\sigma$, and the sum runs over wavevectors
with $k_z>0$, corresponding to forward propagation.
The Hermitian conjugate applies to both terms inside
the brackets.
The energy current is given by the expectation value
$J_i^E=\langle\hat J_i^E\rangle$.
The diagonal contribution describes transport associated
with mode populations and is determined by the eigenfrequencies and group velocities.
The off-diagonal contribution depends on the interband
coherence
$\langle\hat a^\dagger_{\mathbf k,\sigma}
\hat a_{\mathbf k,\sigma'}\rangle$
and combines the frequency splitting with the momentum
variation of the eigenvectors.
Including ferron-photon hybridization, we obtain the longitudinal
and transverse phonon energy currents to  the leading order
in the magnetic field components $B_x$ and $B_z$ as follows, 
\begin{equation}
\begin{aligned}
J^E_{z,k}
&=
\frac{\hbar}{V_c}F_k\omega_k
\frac{\partial\omega_k}{\partial k},
\\
\mathbf J^E_{T,k}
&=
\frac{F_k\mu_{\mathrm{ph}}^2B_xB_z}
     {2V_c\hbar k}
\Bigg[
\frac{h_k^2}{2}\hat{\mathbf x}
\\
&\quad+
h_k\eta(2-\eta)\frac{\omega_k}{\Delta}
\left(
\cos(2\theta)\hat{\mathbf x}
+\sin(2\theta)\hat{\mathbf y}
\right)
\Bigg].
\end{aligned}
\label{eq:energy_currents}
\end{equation}
The branch index $\pm$ is suppressed for clarity, and
$\mathbf J^E_{T,k}=J^E_{x,k}\hat{\mathbf x}
+J^E_{y,k}\hat{\mathbf y}$.
Here, $\mu_{\mathrm{ph}}$ parameterizes the phonon Zeeman
splitting, $\Delta=\Omega_L-\Omega_T$ is the frequency separation
between the longitudinal and transverse optical modes, and
$\eta=\tfrac12\left(
\sqrt{\Omega_L/\Omega_T}
+\sqrt{\Omega_T/\Omega_L}
\right)$ accounts for their unequal frequencies in the
magnetic coupling matrix elements.
The dimensionless factor
\[
h_k
=
\frac{\delta\omega_k^{\mathrm{pol}}}
     {\delta\Omega_T^{\mathrm{ph}}}
=
\left.
\frac{\partial\omega_k}{\partial\Omega_T}
\right|_k
\]
quantifies the polariton renormalization of the ferron Zeeman
splitting and is evaluated from the zero-field dispersion.
The factor $F_k$ weighs the lattice contribution; accordingly,
Eq.~\eqref{eq:energy_currents} describes the lattice part of the polariton energy current.
For propagation along $z$, $B_z$ splits the chiral branches,
while $B_x$ couples them to the longitudinal optical mode.
The first term in $\mathbf J^E_{T,k}$ is the angle-independent
diagonal contribution, whereas the second arises from
interband coherence. The relative phase between the chiral
components, set by the incident polarization angle $\theta$,
produces the $\cos(2\theta)$ and $\sin(2\theta)$ dependence
of this off-diagonal contribution.
The full derivation is presented in the Supplemental Material.


\begin{figure*}[t]
 \centering
\includegraphics[width=\textwidth]{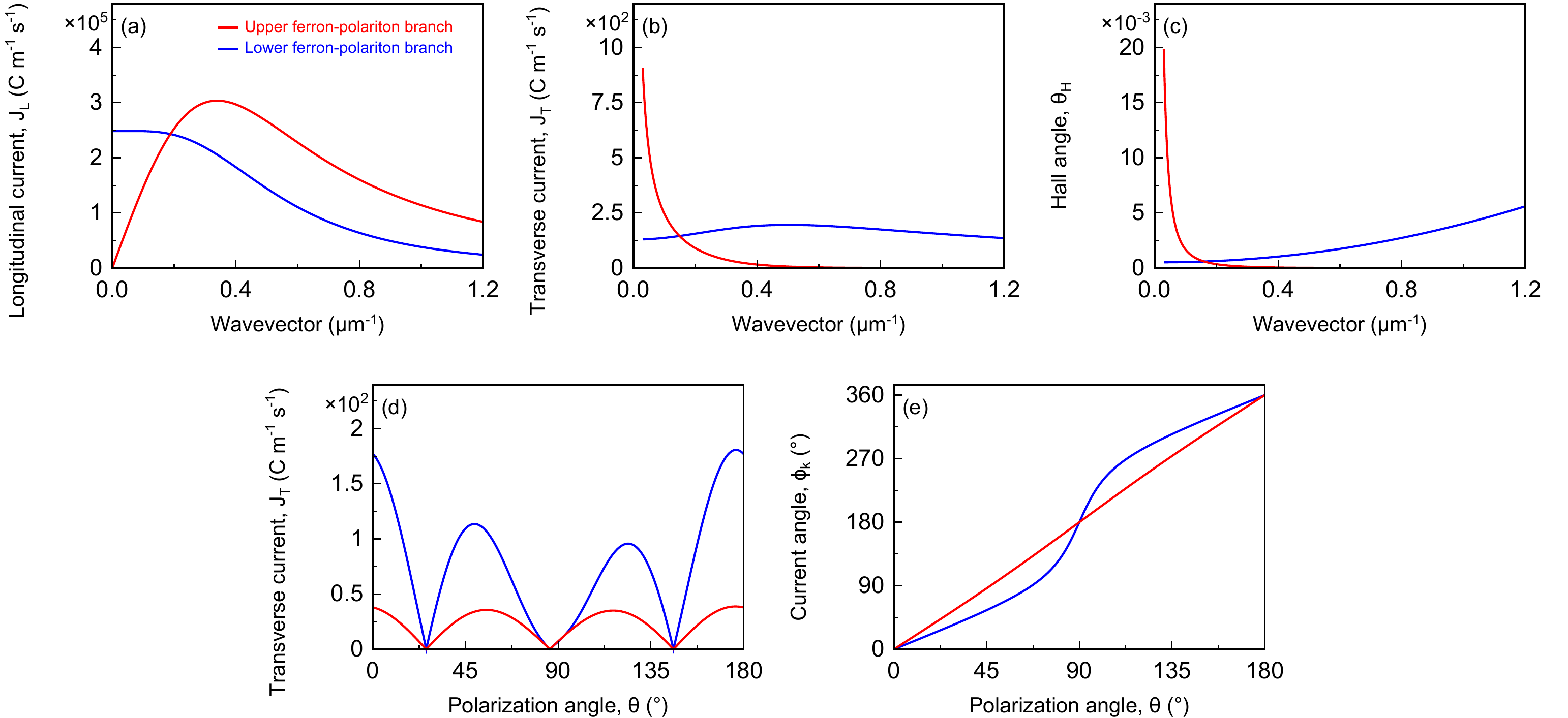}
 \caption{
Magnetic-field-induced net polarization transport and its optical
control in SnTe.
(a-c) Wavevector dependence of the longitudinal polarization
current $J_z^P$, transverse polarization current, 
$J_T^P=\sqrt{(J_x^P)^2+(J_y^P)^2}$, and Hall angle
$\theta_H=J_T^P/J_z^P$, respectively, for incident
pulse polarization along the axis of maximal ferron polarization.
(d,e) Transverse polarization current and angle of the transverse energy flow
$\phi_k$ versus incident pulse polarization angle $\theta$ at
$k=0.3\,\mu\mathrm m^{-1}$, demonstrating optical control of the magnitude of the current and direction of the energy flow.
Blue and red curves denote the lower and upper ferron-polariton branches, respectively.
All panels use $B_z=6\,\mathrm T$, $B_x=1.68\,\mathrm T$,
and $\mu_{\mathrm{ph}}=1\,\mu_B$.
} 
\label{fig:polarization_currents}
\end{figure*} 

Equation~\eqref{eq:energy_currents} gives the transverse current to leading order in $B_x/B_z$, yielding a contribution proportional to $B_xB_z$, under the weak-coupling condition
$|\mu_{\mathrm{ph}}B_z|/\hbar\ll|\Delta|$.
Because higher-order corrections in $B_x/B_z$ are neglected,
the field ratio must also be sufficiently small.
In Supplemental Material, we compare the field-ratio dependence with
numerical calculations using the full magnetic Hamiltonian. For SnTe,
the first omitted correction relative to the leading bilinear
contribution is approximately $0.55(B_x/B_z)^2$ and is therefore
below $5\%$ for $|B_x/B_z|<0.3$.

We obtain the net polarization current by converting the lattice
energy current into transport of the polarization component
$P(\theta)$. Since $V_cP(\theta)$ is the net electric dipole moment
per bare phonon excitation, we can write the polarization current as 
\begin{equation}
J^P_{i,k}
=
\frac{V_cP(\theta)}
     {\hbar\omega_k}
J^E_{i,k}.
\label{eq:polarization_conversion}
\end{equation}
Here, $i=x,y,z$ labels the spatial coordinates.
The transverse polarization current is
$\mathbf J_T^P=J_x^P\hat{\mathbf x}+J_y^P\hat{\mathbf y}$,
with magnitude
$J_T^P=|\mathbf J_T^P|
=\sqrt{(J_x^P)^2+(J_y^P)^2}$
and the Hall angle of $\theta_H=J_T^P/J_z^P$.
For nonzero $P(\theta)$ and longitudinal current, the magnitude of
the conversion prefactor cancels, giving
$\theta_H=J_T^P/J_z^P=J_T^E/J_z^E$, where
$J_T^E=|\mathbf J_T^E|$. We adopt $\mu_{\mathrm{ph}}=1\,\mu_B$, within the experimentally
reported range and below the maximum measured value in
Pb$_{1-x}$Sn$_x$Te, and compute these quantities for
$B_z=6\,\mathrm T$ and $B_x=1.68\,\mathrm T$.
These fields give $|B_x/B_z|=0.28$ and
$|\mu_{\mathrm{ph}}B_z|/(\hbar|\Delta|)\simeq0.043$,
satisfying both validity conditions stated above.

Figure~\ref{fig:polarization_currents}(a) shows the longitudinal
polarization current for the incident pulse polarization specified in the caption.
Its wavevector dependence reflects the competition between
phonon fraction and group velocity: with increasing wavevector,
the lower branch becomes more phonon-like but slower, whereas
the upper branch becomes faster but less phonon-like.
Figure~\ref{fig:polarization_currents}(b) shows the transverse
current. 
The wavevector dependence reflects the phonon fraction, the
renormalization factor $h_k$, and the explicit $1/k$ factor in
Eq.~\eqref{eq:energy_currents}.
The corresponding Hall angle is shown in
Fig.~\ref{fig:polarization_currents}(c), reaching
$\theta_H\sim10^{-3}$ to $10^{-2}$ over the relevant range of propagating ferron polaritons.
An increasing Hall angle can reflect a decreasing longitudinal
current as well as enhanced transverse flow, so the ratio must
be considered together with the absolute current strengths.

We next demonstrate how the transverse current can be controlled through the polarization angle $\theta$ of the incident pulse. We fix
$k=0.3\,\mu\mathrm m^{-1}$ and vary the incident polarization
angle $\theta$.
Figure~\ref{fig:polarization_currents}(d) shows the transverse
polarization current versus $\theta$. Rotating the pulse changes
both the net polarization $P(\theta)$ carried by the excitation and
the transverse energy current in Eq.~\eqref{eq:energy_currents};
their product determines the polarization current through
Eq.~\eqref{eq:polarization_conversion}.
The polarization current vanishes for three distinct angles, for which $P(\theta)=0$, even if the energy current remains finite.
We characterize the direction of transverse energy flow by
the angle $\phi_k=\tan^{-1}(J^E_{y,k}/J^E_{x,k})$,
with the quadrant determined by the signs of the current
components.
Figure~\ref{fig:polarization_currents}(e) shows how this
direction changes with the incident pulse polarization angle $\theta$.
When the off-diagonal contribution dominates over the diagonal
term, the transverse current is approximately proportional to
$B_xB_z[\cos(2\theta)\hat{\mathbf x}
+\sin(2\theta)\hat{\mathbf y}]$.
Thus, for the fields used here, $B_x>0$ and $B_z>0$,
the energy-flow angle approximately follows $\phi_k\simeq2\theta$. Rotating the incident polarization therefore changes the direction of the transverse energy flow. At the same time, the magnetic field provides an independent control: reversing either $B_x$ or $B_z$ reverses the transverse current.


\section{Discussion}

Our results establish a mechanism for deflecting optically rectified polarizations through a ferron polariton Hall
effect. This optically driven response differs  from the recently-proposed thermally driven ferron Hall effect involving acoustic phonons~\cite{lopez2026ferron}, and from the
phonon polariton Hall effect that does not consider transport of net electric polarization~\cite{yaniv2025phonon}. 
The net polarization per phonon scales as $Za/(V_c\omega^3)$
and therefore as $Za/(V_c\omega^4)$ per deposited energy,
making low-frequency optical phonons, such as those in SnTe,
well suited for generating a large ferron polarization.

While we investigated SnTe, the ferron polariton Hall effect is straightforwardly applicable to other ferroelectrics that host $E$- or $T$-symmetric lattice vibrations, which carry a phonon magnetic moment.
Because measurements of the phonon Zeeman effect in ferroelectrics are still scarce, ferroelectrics exhibiting a thermal Hall response provide
a useful starting point for identifying further candidates. Examples of further candidates include multiferroic YMnO$_3$~\cite{kim2024thermal,Oh2024}
and polar (Zn$_x$Fe$_{1-x}$)$_2$Mo$_3$O$_8$~\cite{ideue2017giant},
both of which exhibit phonon thermal Hall responses.

The effect could be probed by tracking a terahertz-excited
ferron polariton wave packet in a magnetic field
using spatially resolved pump-probe
reflectance, with complementary
second-harmonic generation measurements probing the induced
polarization~\cite{Choe2026Ferrons,luo2024time,biggs2026measuring,bakker1998coherent,biggs2026ultrafast}.
Using a bare-phonon lifetime of $\tau_{\mathrm{ph}}\sim5$-$10\,\mathrm{ps}$, estimated from the transverse-optical phonon linewidths reported below the ferroelectric
transition~\cite{o2017inelastic}, and assuming lattice-dominated polariton damping, we obtain $\tau_{\mathrm{pol}}\simeq\tau_{\mathrm{ph}}/F_k$.
This yields an energy attenuation length
$L=v_g\tau_{\mathrm{pol}}\simeq60$-$120\,\mu\mathrm m$
for the lower branch at $k=0.4\,\mu\mathrm m^{-1}$.
At the same wavevector,
Fig.~\ref{fig:polarization_currents}(c) gives
$\theta_H\simeq10^{-3}$, corresponding to a transverse
displacement $\delta r_\perp\simeq\theta_H L
\simeq60$-$120\,\mathrm{nm}$ over one attenuation length.
Comparing trajectories upon reversal of either
magnetic-field component would double their separation and help isolate the Hall response.

Finally, we note that net polarization is carried by a single ferron. An intriguing future direction to explore is therefore whether ferron polaritons could be utilized to achieve single-photon nonlinearity \cite{chang2014quantum,zasedatelev2021single}, which is important for photon-photon quantum gates and
single-photon switching. To follow this direction, a fully quantum mechanical treatment of the ferron polaritons will be necessary.


\section*{Acknowledgments}
We thank Daniel Bustamante Lopez, Pooja Rani, Xiaoyang Zhu  and Mike Pols for useful discussions. This work was supported by the Army Research Office grant No. W911NF-23-1-0243. O.Y. acknowledges support from the Rashba Excellence Award. Calculations were performed on HPC infrastructure at Tel Aviv University.


\bibliography{Bibliography}

@STRING( EL = "Europhys. Lett.")

@STRING{scrmat    = "Scr. Mater."}

@article{juraschek2:2017,
    author = {Juraschek, D. M. and Fechner, M. and Balatsky, A. V. and Spaldin, N. A.},
    journal = {Phys. Rev. Materials},
    pages = {014401},
    title = {{Dynamical multiferroicity}},
    doi = {10.1103/PhysRevMaterials.1.014401},
    volume = {1},
    year = {2017}
,
  url = {https://doi.org/10.1103/PhysRevMaterials.1.014401},

  number = {1}
}

@article{Juraschek2019,
    author = {Juraschek, D. M. and Spaldin, N. A.},
    doi = {10.1103/PhysRevMaterials.3.064405},
    journal = {Phys. Rev. Materials},
    pages = {064405},
    title = {{Orbital magnetic moments of phonons}},
    volume = {3},
    year = {2019}
,
  url = {https://doi.org/10.1103/PhysRevMaterials.3.064405},
}

@article{Grissonnanche2019,
    author = {Grissonnanche, G. and Legros, A. and Badoux, S. and Lefran{\c{c}}ois, E. and Zatko, V. and Lizaire, M. and Lalibert{\'{e}}, F. and Gourgout, A. and Zhou, J. S. and Pyon, S. and Takayama, T. and Takagi, H. and Ono, S. and Doiron-Leyraud, N. and Taillefer, L.},
    doi = {10.1038/s41586-019-1375-0},
    journal = {Nature},
    pages = {376--380},
    title = {{Giant thermal Hall conductivity in the pseudogap phase of cuprate superconductors}},
    volume = {571},
    year = {2019}
,
  url = {https://doi.org/10.1038/s41586-019-1375-0},

  number = {7765}
}

@article{Li2023phononHall,
    author = {Xiaokang Li and Yo Machida and Alaska Subedi and Zengwei Zhu and Liang Li and Kamran Behnia},
    doi = {10.1038/s41467-023-36750-3},
    journal = {Nat. Commun.},
    pages = {1027},
    title = {{The phonon thermal Hall angle in black phosphorus}},
    volume = {14},
    year = {2023}
,
  url = {https://doi.org/10.1038/s41467-023-36750-3},

  number = {1}
}

@article{Flebus2023,
   author = {B. Flebus and A. H. MacDonald},
   doi = {10.1103/PhysRevLett.131.236301},
   issn = {0031-9007},
   issue = {23},
   journal = {Phys. Rev. Lett.},
   month = {12},
   pages = {236301},
   title = {{Phonon Hall Viscosity of Ionic Crystals}},
   volume = {131},
   year = {2023},

  url = {https://doi.org/10.1103/PhysRevLett.131.236301},

  number = {23}
}

@article{Mangeolle2022,
   author = {Léo Mangeolle and Leon Balents and Lucile Savary},
   doi = {10.1103/PhysRevX.12.041031},
   issn = {2160-3308},
   issue = {4},
   journal = {Phys. Rev. X},
   month = {12},
   pages = {041031},
   title = {{Phonon Thermal Hall Conductivity from Scattering with Collective Fluctuations}},
   volume = {12},
   year = {2022},

  url = {https://doi.org/10.1103/PhysRevX.12.041031},
}

@article{Sun2022,
   author = {Xiao-Qi Sun and Jing-Yuan Chen and Steven A. Kivelson},
   doi = {10.1103/PhysRevB.106.144111},
   issn = {2469-9950},
   issue = {14},
   journal = {Phys. Rev. B},
   month = {10},
   pages = {144111},
   title = {Large extrinsic phonon thermal {Hall} effect from resonant scattering},
   volume = {106},
   year = {2022},

  url = {https://doi.org/10.1103/PhysRevB.106.144111},

  number = {14}
}

@article{Chen2024,
author = {Chen, Lu and Lefran{\c{c}}ois, {\'{E}}tienne and Vallipuram, Ashvini and Barth{\'{e}}lemy, Quentin and Ataei, Amirreza and Yao, Weiliang and Li, Yuan and Taillefer, Louis},
doi = {10.1038/s41467-024-47858-5},
journal = {Nat. Commun.},
pages = {3513},
title = {{Planar thermal Hall effect from phonons in a Kitaev candidate material}},
volume = {15},
year = {2024}
,
  url = {https://doi.org/10.1038/s41467-024-47858-5},
}

@article{Jin2024_PHE,
      title={{Discovery of universal phonon thermal Hall effect in crystals}}, 
      author={Xiaobo Jin and Xu Zhang and Wenbo Wan and Hanru Wang and Yihan Jiao and Shiyan Li},
      year={2025},
      journal = {Phys. Rev. Lett.},
      url={https://doi.org/10.1103/r572-5dfm}
,
  volume = {135},
  number = {19},
  pages = {196302},
  doi = {10.1103/r572-5dfm}
}

@article{Oh2024,
author = {Oh, Taekoo and Nagaosa, Naoto},
doi = {10.1103/PhysRevX.15.011036},
journal = {Phys. Rev. X},
pages = {011036},
title = {{Phonon Thermal Hall Effect in Mott Insulators via Skew-Scattering by the Scalar Spin Chirality}},
volume = {15},
year = {2025}
,
  url = {https://doi.org/10.1103/PhysRevX.15.011036},

  number = {1}
}

@article{Guo2021phononHall,
   author = {Haoyu Guo and Subir Sachdev},
   doi = {10.1103/PhysRevB.103.205115},
   issn = {2469-9950},
   issue = {20},
   journal = {Phys. Rev. B},
   month = {5},
   pages = {205115},
   title = {Extrinsic phonon thermal {Hall} transport from {Hall} viscosity},
   volume = {103},
   year = {2021},

  url = {https://doi.org/10.1103/PhysRevB.103.205115},

  number = {20}
}

@article{Grissonnanche2020,
    author = {Grissonnanche, G. and Th{\'{e}}riault, S. and Gourgout, A. and Boulanger, M. E. and Lefran{\c{c}}ois, E. and Ataei, A. and Lalibert{\'{e}}, F. and Dion, M. and Zhou, J. S. and Pyon, S. and Takayama, T. and Takagi, H. and Doiron-Leyraud, N. and Taillefer, L.},
    doi = {10.1038/s41567-020-0965-y},
    journal = {Nat. Phys.},
    pages = {1108--1111},
    title = {{Chiral phonons in the pseudogap phase of cuprates}},
    volume = {16},
    year = {2020}
,
  url = {https://doi.org/10.1038/s41567-020-0965-y},
}

@article{Basini2024,
    author = {Basini, M. and Pancaldi, M. and Wehinger, B. and Udina, M. and Unikandanunni, V. and Tadano, T. and Hoffmann, M. C. and Balatsky, A. V. and Bonetti, S.},
    doi = {10.1038/s41586-024-07175-9},
    journal = {Nature},
    pages = {534--539},
    title = {{Terahertz electric-field driven dynamical multiferroicity in SrTiO$_3$}},
    year = {2024}
,
  url = {https://doi.org/10.1038/s41586-024-07175-9},

  volume = {628},
  number = {8008}
}

@article{Zhang2023_BLG,
    author = {Zhang, X.-W. and Ren, Y. and Wang, C. and Cao, T. and Xiao, D.},
    doi = {10.1103/PhysRevLett.130.226302},
    journal = {Phys. Rev. Lett.},
    pages = {226302},
    title = {{Gate-Tunable Phonon Magnetic Moment in Bilayer Graphene}},
    volume = {130},
    year = {2023}
,
  url = {https://doi.org/10.1103/PhysRevLett.130.226302},
}

@article{Shabala2024,
  title = {{Phonon Inverse Faraday Effect from Electron-Phonon Coupling}},
  author = {Shabala, N. and Geilhufe, R. M.},
  journal = {Phys. Rev. Lett.},
  volume = {133},
  pages = {266702},
  year = {2024},
  doi = {10.1103/PhysRevLett.133.266702}
,
  url = {https://doi.org/10.1103/PhysRevLett.133.266702},

  number = {26}
}

@article{Klebl2024,
  title = {{Ultrafast Pseudomagnetic Fields from Electron-Nuclear Quantum Geometry}},
  author = {Klebl, L. and Schobert, A. and Eckstein, M. and Sangiovanni, G. and Balatsky, A. V. and Wehling, T. O.},
  journal = {Phys. Rev. Lett.},
  volume = {134},
  pages = {016705},
  year = {2025},
  doi = {10.1103/PhysRevLett.134.016705}
,
  url = {https://doi.org/10.1103/PhysRevLett.134.016705},
}

@book{Horton1974,
   author    = {Horton, G. K. and Maradudin, A. A.}, 
   title     = {Dynamical properties of solids},
   publisher = {North-Holland Publishing Company},
   year      = 1974,
   volume    = {1},
   isbn      = {0720402786}
}

@article{strohm:2005,
    author = {Strohm, C. and Rikken, G. L. J. A. and Wyder, P.},
    journal = {Phys. Rev. Lett.},
    doi = {10.1103/PhysRevLett.95.155901},
    title = {{Phenomenological Evidence for the Phonon Hall Effect}},
    volume = {95},
    pages={155901},
    year = {2005}
,
  url = {https://doi.org/10.1103/PhysRevLett.95.155901},

  number = {15}
}

@article{Inyushkin2007,
    author = {Inyushkin, A. V. and Taldenkov, A. N.},
    doi = {10.1134/S0021364007180075},
    journal = {JETP Lett.},
    pages = {379--382},
    title = {{On the Phonon Hall Effect in a Paramagnetic Dielectric}},
    volume = {86},
    year = {2007}
,
  url = {https://doi.org/10.1134/S0021364007180075},

  number = {6}
}

@article{Kagan2008,
    author = {Kagan, Yu. and Maksimov, L. A.},
    doi = {10.1103/PhysRevLett.100.145902},
    journal = {Phys. Rev. Lett.},
    pages = {145902},
    title = {{Anomalous Hall Effect for the Phonon Heat Conductivity in Paramagnetic Dielectrics}},
    volume = {100},
    year = {2008}
,
  url = {https://doi.org/10.1103/PhysRevLett.100.145902},

  number = {14}
}

@article{Wang2009,
    author = {Wang, J. S. and Zhang, L.},
    doi = {10.1103/PhysRevB.80.012301},
    journal = {Phys. Rev. B},
    pages = {012301},
    title = {{Phonon Hall thermal conductivity from the Green-Kubo formula}},
    volume = {80},
    year = {2009}
,
  url = {https://doi.org/10.1103/PhysRevB.80.012301},

  number = {1}
}

@article{Zhang2011,
    author = {Zhang, L. and Ren, J. and Wang, J.-S. and Li, B.},
    doi = {10.1088/0953-8984/23/30/305402},
    journal = {J. Phys.: Condens. Matter},
    pages = {305402},
    title = {{The phonon Hall effect: theory and application}},
    volume = {23},
    year = {2011}
,
  url = {https://doi.org/10.1088/0953-8984/23/30/305402},
}

@article{Qin2012,
    author = {Qin, T. and Zhou, J. and Shi, J.},
    doi = {10.1103/PhysRevB.86.104305},
    pages = {104305},
    journal = {Phys. Rev. B},
    title = {{Berry curvature and the phonon Hall effect}},
    volume = {86},
    year = {2012}
,
  url = {https://doi.org/10.1103/PhysRevB.86.104305},
}

@article{phonopy,
    Author        = {A. Togo and I. Tanaka},
    Journal       = scrmat,
    Title         = {First principles phonon calculations in materials science},
    doi = {10.1016/j.scriptamat.2015.07.021},
    Volume        = {108},
    Pages         = {1--5},
    Year          = {2015}
}

@article{Geilhufe2021,
    author = {Geilhufe, R. M. and Juri{\v{c}}i{\'{c}}, V. and Bonetti, S. and Zhu, J.-X. and Balatsky, A. V.},
    doi = {10.1103/PhysRevResearch.3.L022011},
    journal = {Phys. Rev. Research},
    pages = {L022011},
    title = {{Dynamically induced magnetism in KTaO$_3$}},
    volume = {3},
    year = {2021}
,
  url = {https://doi.org/10.1103/PhysRevResearch.3.L022011},

  number = {2}
}

@article{Li2020,
    author = {Li, Xiaokang and Fauqu{\'{e}}, Beno{\^{i}}t and Zhu, Zengwei and Behnia, Kamran},
    doi = {10.1103/physrevlett.124.105901},
    journal = {Phys. Rev. Lett.},
    pages = {105901},
    title = {{Phonon Thermal Hall Effect in Strontium Titanate}},
    volume = {124},
    year = {2020}
,
  url = {https://doi.org/10.1103/physrevlett.124.105901},
}

@article{Xiao2021,
    author = {Xiao, C. and Ren, Y. and Xiong, B.},
    doi = {10.1103/PhysRevB.103.115432},
    journal = {Phys. Rev. B},
    pages = {115432},
    title = {{Adiabatically induced orbital magnetization}},
    volume = {103},
    year = {2021}
,
  url = {https://doi.org/10.1103/PhysRevB.103.115432},

  number = {11}
}

@article{Ren2021,
    author = {Ren, Y. and Xiao, C. and Saparov, D. and Niu, Q.},
    doi = {10.1103/PhysRevLett.127.186403},
    journal = {Phys. Rev. Lett.},
    pages = {186403},
    title = {{Phonon Magnetic Moment from Electronic Topological Magnetization}},
    volume = {127},
    year = {2021}
,
  url = {https://doi.org/10.1103/PhysRevLett.127.186403},

  number = {18}
}

@article{Baydin2022,
    author = {Baydin, A. and Hernandez, F. G. G. and Rodriguez-Vega, M. and Okazaki, A. K. and Tay, F. and Noe, G. T. and Katayama, I. and Takeda, J. and Nojiri, H. and Rappl, P. H. O. and Abramof, E. and Fiete, G. A. and Kono, J.},
    doi = {10.1103/PhysRevLett.128.075901},
    journal = {Phys. Rev. Lett.},
    pages = {075901},
    title = {{Magnetic Control of Soft Chiral Phonons in PbTe}},
    volume = {128},
    year = {2022}
,
  url = {https://doi.org/10.1103/PhysRevLett.128.075901},

  number = {7}
}

@article{Mustafa2025,
    author = {Mustafa, Hussam and Nnokwe, Cynthia and Ye, Gaihua and Fang, Mengqi and Chaudhary, Swati and Yan, Jia An and Wu, Kai and Cunningham, Connor J. and Hemesath, Colin M. and Stollenwerk, Andrew James and Shand, Paul M. and Yang, Eui Hyeok and Fiete, Gregory A. and He, Rui and Jin, Wencan},
    doi = {10.1021/acsnano.4c18906},
    journal = {ACS Nano},
    pages = {11241--11248},
    title = {{Origin of Large Effective Phonon Magnetic Moments in Monolayer MoS$_2$}},
    volume = {19},
    year = {2025}
,
  url = {https://doi.org/10.1021/acsnano.4c18906},

  number = {11}
}

@article{Paiva2025,
  title = {{Dynamically Induced Multiferroic Polarization}},
  author = {Paiva, Carolina and Fechner, Michael and Juraschek, Dominik M.},
  journal = {Phys. Rev. Lett.},
  volume = {135},
  pages = {066702},
  year = {2025},
  doi = {10.1103/7lm1-wm3y}
,
  url = {https://doi.org/10.1103/7lm1-wm3y},

  number = {6}
}

@article{Chen2025_gaugetheory,
  title = {{Gauge theory of giant phonon magnetic moment in doped Dirac semimetals}},
  author = {Chen, W. and Zhang, X.-W. and Su, Y. and Cao, T. and Xiao, D. and Lin, S.-Z.},
  journal = {Phys. Rev. B},
  volume = {111},
  pages = {035126},
  year = {2025},
  doi = {10.1103/PhysRevB.111.035126}
,
  url = {https://doi.org/10.1103/PhysRevB.111.035126},
}

@article{Cheng2020,
    author = {Cheng, B. and Schumann, T. and Wang, Y. and Zhang, X. and Barbalas, D. and Stemmer, S. and Armitage, N. P.},
    doi = {10.1021/acs.nanolett.0c01983},
    journal = {Nano Lett.},
    pages = {5991--5996},
    title = {{A Large Effective Phonon Magnetic Moment in a Dirac Semimetal}},
    volume = {20},
    year = {2020}
,
  url = {https://doi.org/10.1021/acs.nanolett.0c01983},

  number = {8}
}

@article{Geilhufe2023,
    author = {Geilhufe, R. M. and Hergert, W.},
    doi = {10.1103/PhysRevB.107.L020406},
    journal = {Phys. Rev. B},
    title = {{Electron magnetic moment of transient chiral phonons in KTaO$_3$}},
    pages = {L020406},
    volume = {107},
    year = {2023}
,
  url = {https://doi.org/10.1103/PhysRevB.107.L020406},

  number = {2}
}

@article{Hernandez2023,
    author = {Hernandez, F. G.G. and Baydin, A. and Chaudhary, S. and Tay, F. and Katayama, I. and Takeda, J. and Nojiri, H. and Okazaki, A. K. and Rappl, P. H. O. and Abramof, E. and Rodriguez-Vega, M. and Fiete, G. A. and Kono, J.},
    doi = {10.1126/sciadv.adj4074},
    journal = {Sci. Adv.},
    pages = {eadj4074},
    title = {{Observation of interplay between phonon chirality and electronic band topology}},
    volume = {9},
    year = {2023}
,
  url = {https://doi.org/10.1126/sciadv.adj4074},

  number = {50}
}

@article{luo2024time,
  title={Time-of-flight detection of terahertz phonon-polariton},
  author={Luo, Tianchuang and Ilyas, Batyr and Hoegen, A von and Lee, Youjin and Park, Jaena and Park, Je-Geun and Gedik, Nuh},
  doi={10.1038/s41467-024-46515-1},
  journal={Nat. Commun.},
  volume={15},
  pages={2276},
  year={2024}
,
  url = {https://doi.org/10.1038/s41467-024-46515-1},
}

@article{sheng2006theory,
  title={Theory of the phonon {Hall} effect in paramagnetic dielectrics},
  author={Sheng, L and Sheng, D.~N. and Ting, C.~S.},
  doi={10.1103/PhysRevLett.96.155901},  
  journal={Phys. Rev. Lett.},
  volume={96},
  number={15},
  pages={155901},
  year={2006},
  publisher={APS}
,
  url = {https://doi.org/10.1103/PhysRevLett.96.155901},
}

@article{zhang2010topological,
  title={{Topological Nature of the Phonon Hall Effect}},
  author={Zhang, Lifa and Ren, Jie and Wang, Jian-Sheng and Li, Baowen},
  doi={10.1103/PhysRevLett.105.225901},
  journal={Phys. Rev. Lett.},
  volume={105},
  number={22},
  pages={225901},
  year={2010},
  publisher={APS}
,
  url = {https://doi.org/10.1103/PhysRevLett.105.225901},
}

@article{ideue2017giant,
  title={Giant thermal {Hall} effect in multiferroics},
  author={Ideue, Toshiya and Kurumaji, Takashi and Ishiwata, Shintaro and Tokura, Yoshinori},
  doi={10.1038/NMAT4905},
  journal={Nat. Mater.},
  volume={16},
  number={8},
  pages={797--802},
  year={2017},
  publisher={Nature Publishing Group UK London}
,
  url = {https://doi.org/10.1038/NMAT4905},
}

@article{chen2022large,
  title={{Large phonon thermal Hall conductivity in the antiferromagnetic insulator Cu$_3$TeO$_6$}},
  author={Chen, Lu and Boulanger, Marie-Eve and Wang, Zhi-Cheng and Tafti, Fazel and Taillefer, Louis},
  doi={10.1073/pnas.2208016119},
  journal={Proc. Natl. Acad. Sci. USA},
  volume={119},
  pages={e2208016119},
  year={2022}
,
  url = {https://doi.org/10.1073/pnas.2208016119},

  number = {34}
}

@article{mori2014origin,
  title={{Origin of the Phonon Hall Effect in Rare-Earth Garnets}},
  author={Mori, Michiyasu and Spencer-Smith, Alexander and Sushkov, Oleg P and Maekawa, Sadamichi},
  doi={10.1103/PhysRevLett.113.265901},
  journal={Phys. Rev. Lett.},
  volume={113},
  number={26},
  pages={265901},
  year={2014},
  publisher={APS}
,
  url = {https://doi.org/10.1103/PhysRevLett.113.265901},
}

@article{agarwalla2011phonon,
  title={Phonon {Hall} effect in ionic crystals in the presence of static magnetic field},
  author={Agarwalla, Bijay K and Zhang, Lifa and Wang, Jian-Sheng and Li, Baowen},
  doi={10.1140/epjb/e2011-11002-x},
  journal={Eur. Phys. J. B},
  volume={81},
  number={2},
  pages={197--202},
  year={2011},
  publisher={Springer}
,
  url = {https://doi.org/10.1140/epjb/e2011-11002-x},
}

@article{yang2020universal,
  title={{Universal Behavior of The Thermal Hall Conductivity}},
  author={Yang, Yi-feng and Zhang, Guang-Ming and Zhang, Fu-Chun},
  doi={10.1103/PhysRevLett.124.186602},
  journal={Phys. Rev. Lett.},
  volume={124},
  number={18},
  pages={186602},
  year={2020},
  publisher={APS}
,
  url = {https://doi.org/10.1103/PhysRevLett.124.186602},
}

@article{Chen2024_PHE,
  title = {Planar Thermal {Hall} Effect from Phonons in Cuprates},
  author = {Chen, Lu and Le Roux, L\'ena and Grissonnanche, Ga\"el and Boulanger, Marie-Eve and Th\'eriault, Steven and Liang, Ruixing and Bonn, D. A. and Hardy, W. N. and Pyon, S. and Takayama, T. and Takagi, H. and Xu, Ke-Jun and Shen, Zhi-Xun and Taillefer, Louis},
  journal = {Phys. Rev. X},
  volume = {14},
  pages = {041011},
  year = {2024},
  doi = {10.1103/PhysRevX.14.041011}
,
  url = {https://doi.org/10.1103/PhysRevX.14.041011},

  number = {4}
}

@article{Juraschek2025,
author={Juraschek, Dominik M.
and Geilhufe, R. Matthias
and Zhu, Hanyu
and Basini, Martina
and Baum, Peter
and Baydin, Andrey
and Chaudhary, Swati
and Fechner, Michael
and Flebus, Benedetta
and Grissonnanche, Gael
and Kirilyuk, Andrei I.
and Lemeshko, Mikhail
and Maehrlein, Sebastian F.
and Mignolet, Maxime
and Murakami, Shuichi
and Niu, Qian
and Nowak, Ulrich
and Romao, Carl P.
and Rostami, Habib
and Satoh, Takuya
and Spaldin, Nicola A.
and Ueda, Hiroki
and Zhang, Lifa},
title={Chiral phonons},
journal={Nat. Phys.},
year={2025},
volume={21},
pages={1532--1540},
doi={10.1038/s41567-025-03001-9}
,
  url = {https://doi.org/10.1038/s41567-025-03001-9},

  number = {10}
}

@article{Flebus2022,
   author = {B. Flebus and A. H. MacDonald},
   doi = {10.1103/PhysRevB.105.L220301},
   issn = {2469-9950},
   issue = {22},
   journal = {Phys. Rev. B},
   month = {6},
   pages = {L220301},
   title = {Charged defects and phonon {Hall} effects in ionic crystals},
   volume = {105},
   year = {2022}
,
  url = {https://doi.org/10.1103/PhysRevB.105.L220301},

  number = {22}
}

@article{pols2026lattice,
  title={Lattice Excitations with Finite Polarization and Magnetization},
  author={Pols, Mike and Romao, Carl P and Juraschek, Dominik M},
  journal={Phys. Rev. Lett.},
  volume={136},
  number={16},
  pages={166801},
  year={2026},
  publisher={APS}
,
  doi = {10.1103/2sj3-33ky},
  url = {https://doi.org/10.1103/2sj3-33ky},
}

@article{yaniv2025phonon,
  title={Phonon polariton {Hall} effect},
  author={Yaniv, Omer and Juraschek, Dominik M},
  journal={arXiv:2509.16100},
  year={2025}
,
  url = {https://arxiv.org/abs/2509.16100},
}

@article{plekhanov2014engineering,
  title={Engineering relativistic effects in ferroelectric {SnTe}},
  author={Plekhanov, Evgeny and Barone, Paolo and Di Sante, Domenico and Picozzi, Silvia},
  journal={Phys. Rev. B},
  year={2014}
,
  url = {https://doi.org/10.1103/PhysRevB.90.161108},

  volume = {90},
  number = {16},
  pages = {161108},
  doi = {10.1103/PhysRevB.90.161108}
}

@article{bauerPolarizationTransportFerroelectrics2023,
  title = {Polarization Transport in Ferroelectrics},
  author = {Bauer, G.E.W. and Tang, P. and Iguchi, R. and Xiao, J. and Shen, K. and Zhong, Z. and Yu, T. and Rezende, S.M. and Heremans, J.P. and Uchida, K.},
  year = {2023},
  month = nov,
  journal = {Phys. Rev. Applied},
  volume = {20},
  number = {5},
  pages = {050501},
  publisher = {American Physical Society},
  doi = {10.1103/PhysRevApplied.20.050501},

  url = {https://doi.org/10.1103/PhysRevApplied.20.050501},
}

@article{tangElectricAnalogMagnons2024,
  title = {Electric Analog of Magnons in Order-Disorder Ferroelectrics},
  author = {Tang, Ping and Bauer, Gerrit E. W.},
  year = {2024},
  month = feb,
  journal = {Phys. Rev. B},
  volume = {109},
  number = {6},
  pages = {L060301},
  publisher = {American Physical Society},
  doi = {10.1103/PhysRevB.109.L060301},

  url = {https://doi.org/10.1103/PhysRevB.109.L060301},
}

@article{zhouSurfaceFerronExcitations2023,
  title = {Surface {{Ferron Excitations}} in {{Ferroelectrics}} and {{Their Directional Routing}}},
  author = {Zhou, Xi-Han and Cai, Chengyuan and Tang, Ping and {Rodr{\'i}guez-Su{\'a}rez}, R. L. and Rezende, Sergio M. and Bauer, Gerrit E. W. and Yu, Tao},
  year = {2023},
  month = aug,
  journal = {Chin. Phys. Lett.},
  volume = {40},
  number = {8},
  pages = {087103},
  publisher = {{Chinese Physical Society and IOP Publishing Ltd}},
  issn = {0256-307X},
  doi = {10.1088/0256-307X/40/8/087103},
  url = {https://doi.org/10.1088/0256-307X/40/8/087103}
}

@article{morozovskaFlexocouplinginducedPhononsFerrons2025,
  title = {Flexocoupling-Induced Phonons and Ferrons in van Der {{Waals}} Ferroelectrics},
  author = {Morozovska, Anna N. and Eliseev, Eugene A. and Bereznikov, Oleksiy V. and Yelisieiev, Mykola Ye. and Zhao, Guo-Dong and Zhu, Yujie and Gopalan, Venkatraman and Chen, Long-Qing and Hu, Jia-Mian and Vysochanskii, Yulian M.},
  year = {2025},
  month = jul,
  journal = {Phys. Rev. B},
  volume = {112},
  number = {1},
  pages = {014110},
  publisher = {American Physical Society},
  doi = {10.1103/mr9v-khtj},

  url = {https://doi.org/10.1103/mr9v-khtj},
}

@article{zhaoRoleFerronsHeat2025,
  title = {Role of Ferrons in the Heat Capacity and Thermal Transport of Displacive Ferroelectrics},
  author = {Zhao, G. D. and Yang, F. and Chen, L. Q.},
  year = {2025},
  month = jul,
  journal = {Phys. Rev. B},
  volume = {112},
  number = {1},
  pages = {014115},
  publisher = {American Physical Society},
  doi = {10.1103/3y1m-66s1},

  url = {https://doi.org/10.1103/3y1m-66s1},
}

@misc{shenObservationFerronTransport2025,
  title = {Observation of Ferron Transport in Ferroelectrics},
  author = {Shen, Kaiwen and Tang, Ping and Chen, Xianzhe and Gao, Yifan and Fan, Yuanfei and Guo, Zejing and Wei, Yingfen and Jiang, Hao and Zhang, Xumeng and Wang, Ming and He, Pan and Shi, Wu and Han, Jiahao and Wu, Yizheng and Shen, Jian and Liu, Qi and Bauer, Gerrit E. W. and Liu, Ming},
  year = {2025},
  month = may,
  number = {arXiv:2505.24419},
  eprint = {2505.24419},
  primaryclass = {physics},
  publisher = {arXiv},
  doi = {10.48550/arXiv.2505.24419},

  url = {https://doi.org/10.48550/arXiv.2505.24419},
}

@article{Subedi2026Ferrons,
      title={Electrically switchable ferron upconversion in a van der {Waals} ferroelectric}, 
      author={Sujan Subedi and Wuzhang Fang and Fan Fei and Zixin Zhai and Jack P. Rollins and Carter Fox and Alaina Drew and Bing Lv and Yuan Ping and Jun Xiao},
      year={2026},
      journal={arXiv:2603.19394},
      url={https://arxiv.org/abs/2603.19394}
}

@Article{Jana2026Ferrons,
author={Jana, Sambhu
and Zhang, Baolong
and Mishra, Sobhan Subhra
and Tan, Thomas CaiWei
and Chaturvedi, Apoorva
and Teo, Edwin Hang Tong
and Singh, Ranjan},
  title   = {{Ferron-driven photoferroic hysteresis in van der Waals CuInP$_2$S$_6$}},
  journal = {Nat. Commun.},
  year    = {2026},
  volume  = {17},
  number  = {1},
  pages   = {7100},
  doi     = {10.1038/s41467-026-73771-0},
  url     = {https://doi.org/10.1038/s41467-026-73771-0}
}

@Article{Choe2026Ferrons,
author={Choe, Jeongheon
and Handa, Taketo
and Huang, Chun-Ying
and Liston, Andr{\'e} Koch
and Cox, Jordan
and Stensberg, Jonathan
and Hong, Yongseok
and Chica, Daniel G.
and Xu, Ding
and Tay, Fuyang
and Husremovic, Samra
and da Silveira Lanza Avelar, Vinicius
and Arsenault, Eric A.
and Zhang, Zhuquan
and McIver, James
and Basov, Dmitri N.
and Delor, Milan
and Roy, Xavier
and Zhu, X.-Y.},
title={Observation of coherent ferron emission and propagation},
journal={Nat. Mater.},
year={2026},
doi={10.1038/s41563-026-02597-4}
,
  url = {https://doi.org/10.1038/s41563-026-02597-4},

  volume = {25},
  number = {10},
  pages = {1755--1761}
}

@article{zhang2026electric,
  title={Electric-field control of coherent ferron oscillations},
  author={Zhang, Baolong and Duan, Ruihuan and Mishra, Sobhan Subhra and Jana, Sambhu and Kim, Jonghyeon and Tan Caiwei, Thomas and Tan, Yi Ji and Wang, Wenhao and Chen Ietro, Pang Teng and Liu, Zheng and Singh, Ranjan},
  journal={Nat. Phys.},
  volume={22},
  number={10},
  
  year={2026},
  publisher={Nature Publishing Group UK London}
,
  doi = {10.1038/s41567-026-03406-0},
  url = {https://doi.org/10.1038/s41567-026-03406-0},
}

@article{tang2022excitations,
  title={Excitations of the ferroelectric order},
  author={Tang, Ping and Iguchi, Ryo and Uchida, Ken-ichi and Bauer, Gerrit EW},
  journal={Phys. Rev. B},
  volume={106},
  number={8},
  pages={L081105},
  year={2022},
  publisher={APS}
,
  doi = {10.1103/PhysRevB.106.L081105},
  url = {https://doi.org/10.1103/PhysRevB.106.L081105},
}

@article{bauer2022magnonics,
  title={Magnonics vs. ferronics},
  author={Bauer, Gerrit EW and Tang, Ping and Iguchi, Ryo and Uchida, Ken-ichi},
  journal={J. Magn. Magn. Mater.},
  volume={541},
  pages={168468},
  year={2022},
  publisher={Elsevier}
,
  doi = {10.1016/j.jmmm.2021.168468},
  url = {https://doi.org/10.1016/j.jmmm.2021.168468},
}

@article{kim2024thermal,
  title={Thermal {Hall} effects due to topological spin fluctuations in {YMnO$_3$}},
  author={Kim, Ha-Leem and Saito, Takuma and Yang, Heejun and Ishizuka, Hiroaki and Coak, Matthew John and Lee, Jun Han and Sim, Hasung and Oh, Yoon Seok and Nagaosa, Naoto and Park, Je-Geun},
  journal={Nat. Commun.},
  volume={15},
  number={1},
  pages={243},
  year={2024},
  publisher={Nature Publishing Group UK London}
,
  doi = {10.1038/s41467-023-44448-9},
  url = {https://doi.org/10.1038/s41467-023-44448-9},
}

@article{tang2026multiferroic,
  title={Multiferroic-like quasiparticles in ferroelectrics},
  author={Tang, Ping and Bauer, Gerrit EW},
  journal={Phys. Rev. Lett.},
  volume={136},
  number={17},
  pages={176702},
  year={2026},
  publisher={APS}
,
  doi = {10.1103/gm7n-8ftp},
  url = {https://doi.org/10.1103/gm7n-8ftp},
}

@article{lopez2026ferron,
  title={Ferron {Hall} effect: Transverse accumulation of polarization driven by thermal gradients in ferroelectrics},
  author={Lopez, Daniel A Bustamante and Brehm, Verena and Juraschek, Dominik M},
  journal={arXiv:2606.29765},
  year={2026}
,
  url = {https://arxiv.org/abs/2606.29765},
}

@article{o2017inelastic,
  title={Inelastic x-ray investigation of the ferroelectric transition in {SnTe}},
  author={O'Neill, Christopher D and Sokolov, Dmitry A and Hermann, Andreas and Bossak, Alexei and Stock, Christopher and Huxley, Andrew D},
  journal={Phys. Rev. B},
  volume={95},
  number={14},
  pages={144101},
  year={2017},
  publisher={APS}
,
  doi = {10.1103/PhysRevB.95.144101},
  url = {https://doi.org/10.1103/PhysRevB.95.144101},
}

@article{chang2014quantum,
  title={Quantum nonlinear optics—photon by photon},
  author={Chang, Darrick E and Vuleti{\'c}, Vladan and Lukin, Mikhail D},
  journal={Nat. Photonics},
  volume={8},
  number={9},
  pages={685--694},
  year={2014},
  publisher={Nature Publishing Group UK London}
,
  doi = {10.1038/nphoton.2014.192},
  url = {https://doi.org/10.1038/nphoton.2014.192},
}

@article{zasedatelev2021single,
  title={Single-photon nonlinearity at room temperature},
  author={Zasedatelev, Anton V and Baranikov, Anton V and Sannikov, Denis and Urbonas, Darius and Scafirimuto, Fabio and Shishkov, Vladislav Yu and Andrianov, Evgeny S and Lozovik, Yurii E and Scherf, Ullrich and St{\"o}ferle, Thilo and Mahrt, Rainer F. and Lagoudakis, Pavlos G.},
  journal={Nature},
  volume={597},
  number={7877},
  pages={493--497},
  year={2021},
  publisher={Nature Publishing Group UK London}
,
  doi = {10.1038/s41586-021-03866-9},
  url = {https://doi.org/10.1038/s41586-021-03866-9},
}

@article{biggs2026measuring,
  title={Measuring momentum-resolved dissipation of phonon-polaritons in {LiNbO$_3$} with terahertz driving},
  author={Biggs, Megan F and Acampora, Rossella and Sellati, Niccol{\`o} and Udina, Mattia and Benfatto, Lara and Abreu, Elsa and Johnson, Steven L and Johnson, Jeremy A},
  journal={arXiv:2607.16718},
  year={2026}
  ,
  url = {https://arxiv.org/abs/2607.16718},
}

@article{chumak2015magnon,
  title={Magnon spintronics},
  author={Chumak, Andrii V and Vasyuchka, Vitaliy I and Serga, Alexander A and Hillebrands, Burkard},
  journal={Nat. Phys.},
  volume={11},
  number={6},
  pages={453--461},
  year={2015},
  publisher={Nature Publishing Group UK London}
,
  doi = {10.1038/nphys3347},
  url = {https://doi.org/10.1038/nphys3347},
}

@article{pirro2021advances,
  title={Advances in coherent magnonics},
  author={Pirro, Philipp and Vasyuchka, Vitaliy I and Serga, Alexander A and Hillebrands, Burkard},
  journal={Nat. Rev. Mater.},
  volume={6},
  number={12},
  pages={1114--1135},
  year={2021},
  publisher={Nature Publishing Group UK London}
,
  doi = {10.1038/s41578-021-00332-w},
  url = {https://doi.org/10.1038/s41578-021-00332-w},
}

@article{bakker1998coherent,
  title={Coherent phonon polaritons as probes of anharmonic phonons in ferroelectrics},
  author={Bakker, Huib J and Hunsche, Stefan and Kurz, Heinrich},
  journal={Rev. Mod. Phys.},
  volume={70},
  number={2},
  pages={523--536},
  year={1998},
  publisher={APS}
,
  doi = {10.1103/RevModPhys.70.523},
  url = {https://doi.org/10.1103/RevModPhys.70.523},
}

@article{biggs2026ultrafast,
  title={Ultrafast {Faraday} rotation probe of chiral phonon-polaritons in {LiNbO$_3$}},
  author={Biggs, Megan F and Ho, Sin-hang and Alejandro, Aldair and Lutz, Matthew J and Moss, Clayton D and Johnson, Jeremy A},
  journal={Sci. Adv.},
  volume={12},
  number={36},
  pages={eaec8970},
  year={2026},
  publisher={American Association for the Advancement of Science}
,
  doi = {10.1126/sciadv.aec8970},
  url = {https://doi.org/10.1126/sciadv.aec8970},
}

@article{Kresse1996PRB,
  author  = {Kresse, Georg and Furthm{\"u}ller, J{\"u}rgen},
  title   = {Efficient iterative schemes for ab initio total-energy
             calculations using a plane-wave basis set},
  journal = {Phys. Rev. B},
  volume  = {54},
  number  = {16},
  pages   = {11169--11186},
  year    = {1996},
  doi     = {10.1103/PhysRevB.54.11169}
,
  url = {https://doi.org/10.1103/PhysRevB.54.11169},
}

@article{Kresse1996CMS,
  author  = {Kresse, Georg and Furthm{\"u}ller, J{\"u}rgen},
  title   = {Efficiency of ab-initio total energy calculations for
             metals and semiconductors using a plane-wave basis set},
  journal = {Comput. Mater. Sci.},
  volume  = {6},
  number  = {1},
  pages   = {15--50},
  year    = {1996},
  doi     = {10.1016/0927-0256(96)00008-0}
,
  url = {https://doi.org/10.1016/0927-0256(96)00008-0},
}

@article{Perdew1996PBE,
  author  = {Perdew, John P. and Burke, Kieron and Ernzerhof, Matthias},
  title   = {Generalized Gradient Approximation Made Simple},
  journal = {Phys. Rev. Lett.},
  volume  = {77},
  number  = {18},
  pages   = {3865--3868},
  year    = {1996},
  doi     = {10.1103/PhysRevLett.77.3865}
,
  url = {https://doi.org/10.1103/PhysRevLett.77.3865},
}

@article{togo2023first,
  title={First-principles phonon calculations with {Phonopy} and {Phono3py}},
  author={Togo, Atsushi},
  journal={J. Phys. Soc. Jpn.},
  volume={92},
  number={1},
  pages={012001},
  year={2023},
  publisher={The Physical Society of Japan}
,
  doi = {10.7566/JPSJ.92.012001},
  url = {https://doi.org/10.7566/JPSJ.92.012001},
}

@article{Flebus2024MagnonicsRoadmap,
  author = {Flebus, Benedetta
    and Grundler, Dirk
    and Rana, Bivas
    and Otani, YoshiChika
    and Barsukov, Igor
    and Barman, Anjan
    and Gubbiotti, Gianluca
    and Landeros, Pedro
    and Akerman, Johan
    and Ebels, Ursula
    and Pirro, Philipp
    and Demidov, Vladislav E.
    and Schultheiss, Katrin
    and Csaba, Gyorgy
    and Wang, Qi
    and Ciubotaru, Florin
    and Nikonov, Dmitri E.
    and Che, Ping
    and Hertel, Riccardo
    and Ono, Teruo
    and Afanasiev, Dmytro
    and Mentink, Johan
    and Rasing, Theo
    and Hillebrands, Burkard
    and Kusminskiy, Silvia Viola
    and Zhang, Wei
    and Du, Chunhui Rita
    and Finco, Aurore
    and van der Sar, Toeno
    and Luo, Yunqiu Kelly
    and Shiota, Yoichi
    and Sklenar, Joseph
    and Yu, Tao
    and Rao, Jinwei},
  title   = {The 2024 magnonics roadmap},
  journal = {J. Phys.: Condens. Matter},
  volume  = {36},
  number  = {36},
  pages   = {363501},
  year    = {2024},
  doi     = {10.1088/1361-648X/ad399c},
  url     = {https://doi.org/10.1088/1361-648X/ad399c}
  }

@article{Nursagatov2026FerronPolaritons,
  author  = {Nursagatov, M. and Ye, Xiyin
             and Bobkov, G. A. and Yu, Tao
             and Bobkova, I. V.},
  title   = {Ferron-polaritons in superconductor/ferroelectric/superconductor heterostructures},
  journal = {arXiv:2602.05473},
  year    = {2026},
  url     = {https://arxiv.org/abs/2602.05473}
}

\onecolumngrid
\clearpage

\setcounter{page}{1}

\begin{center}
\textbf{\large Supplemental Material for \\Ferron polariton Hall effect: Deflecting optically rectified polarization with magnetic fields}\\[0.4cm]

Omer Yaniv$^{1}$ and Dominik M. Juraschek$^{1,2,3,4}$\\[0.2cm]

\affiliation{}

$^{1}${\itshape{\small School of Physics and Astronomy, Tel Aviv University,
Tel Aviv 6997801, Israel}}\\
$^{2}${\itshape{\small Department of Applied Physics and Science Education,
Eindhoven University of Technology, 5612 AP Eindhoven, Netherlands}}\\
$^{3}${\itshape{\small Department of Physics, University of Fribourg,
CH-1700 Fribourg, Switzerland}}\\
$^{4}${\itshape{\small Paul Scherrer Institut, CH-5232 Villigen PSI, Switzerland}}
\end{center}

\setcounter{equation}{0}
\setcounter{figure}{0}
\setcounter{table}{0}
\makeatletter
\renewcommand{\theequation}{S\arabic{equation}}
\renewcommand{\thefigure}{S\arabic{figure}}
\renewcommand{\thetable}{S\arabic{table}}

\tableofcontents

\newpage

\section{Transverse interband energy current}

We begin from the form of the interband energy current used directly
in the numerical calculation. For two phonon branches labeled $1$ and
$2$, the time-averaged interband contribution is

\begin{equation}
\label{eq:J-D}
\left\langle J_i^{\mathrm{inter}}\right\rangle
=
\frac{\hbar}{V_c}
\sum_{\mathbf{k}>0}
\mathrm{Re}\!\left[
\bra{\boldsymbol\epsilon_{\mathbf{k},1}}
\partial_{k_i}D(\mathbf{k},\mathbf B)
\ket{\boldsymbol\epsilon_{\mathbf{k},2}}
\;
\left\langle
a_{\mathbf{k},1}^{\dagger}a_{\mathbf{k},2}
\right\rangle
\right].
\end{equation}

Here $D(\mathbf{k},\mathbf B)$ is the magnetic dynamical matrix,
$V_c$ is the unit-cell volume,
$\boldsymbol\epsilon_{\mathbf{k},n}$ are its normalized eigenvectors,
and
$\langle a_{\mathbf{k},1}^{\dagger}a_{\mathbf{k},2}\rangle$
is the interband coherence generated by the coherent drive.

The phonon eigenvalue equation is

$$
D(\mathbf{k},\mathbf B)
\ket{\boldsymbol\epsilon_{\mathbf{k},n}}
=
\omega_{\mathbf{k},n}^{2}
\ket{\boldsymbol\epsilon_{\mathbf{k},n}}.
$$

Equation~(\ref{eq:J-D}) is the form evaluated directly in the numerical
calculation. To obtain an analytical interpretation of this result,
we first rewrite the current in terms of the momentum derivative of
the phonon eigenvectors.

Differentiating the eigenvalue equation for branch $2$ gives

$$
(\partial_{k_i}D)
\ket{\boldsymbol\epsilon_2}
+
D
\ket{\partial_{k_i}\boldsymbol\epsilon_2}
=
(\partial_{k_i}\omega_2^2)
\ket{\boldsymbol\epsilon_2}
+
\omega_2^2
\ket{\partial_{k_i}\boldsymbol\epsilon_2}.
$$

Multiplying from the left by
$\bra{\boldsymbol\epsilon_1}$ and using

$$
\braket{\boldsymbol\epsilon_1|\boldsymbol\epsilon_2}=0,
$$

together with

$$
\bra{\boldsymbol\epsilon_1}D
=
\omega_1^2
\bra{\boldsymbol\epsilon_1},
$$

gives

\begin{equation}
\label{eq:HF}
\bra{\boldsymbol\epsilon_1}
\partial_{k_i}D
\ket{\boldsymbol\epsilon_2}
=
\left(
\omega_2^2-\omega_1^2
\right)
\braket{
\boldsymbol\epsilon_1|
\partial_{k_i}\boldsymbol\epsilon_2
}.
\end{equation}

The current can therefore equivalently be written as

\begin{equation}
\label{eq:J-eigenvector}
\left\langle J_i^{\mathrm{inter}}\right\rangle
=
\frac{\hbar}{V_c}
\sum_{\mathbf{k}>0}
\left(
\omega_{\mathbf{k},2}^2-\omega_{\mathbf{k},1}^2
\right)
\mathrm{Re}\!\left[
\boldsymbol\epsilon_{\mathbf{k},1}^{*}
\cdot
\partial_{k_i}
\boldsymbol\epsilon_{\mathbf{k},2}
\;
\left\langle
a_{\mathbf{k},1}^{\dagger}a_{\mathbf{k},2}
\right\rangle
\right].
\end{equation}

We define

$$
\mathcal A_i
=
\braket{
\boldsymbol\epsilon_{\mathbf{k},1}|
\partial_{k_i}
\boldsymbol\epsilon_{\mathbf{k},2}
}.
$$

The purpose of the following derivation is to calculate
$\mathcal A_x$ analytically within a three-mode model and compare the
resulting field dependence with the current obtained directly from
Eq.~(\ref{eq:J-D}).

\section{Three-mode magnetic dynamical matrix}

We consider propagation along

$$
\mathbf k=k_z\hat{\mathbf z},
$$

and retain the two circular transverse modes $|L\rangle$ and
$|R\rangle$ together with an auxiliary optical mode $|A\rangle$.

At the reference propagation direction, the magnetic three-mode
problem is described by

\begin{equation}
\label{eq:D3}
D_{\mathrm{eff}}
=
\begin{pmatrix}
\Omega_T+a & 0 & -v\\
0 & \Omega_T-a & v\\
-v & v & \Omega_A
\end{pmatrix},
\qquad
a=\gamma B_z,
\qquad
v=\frac{\gamma\eta}{\sqrt2}B_x.
\end{equation}

Here $\Omega_T$ is the frequency of the transverse doublet,
$\Omega_A$ is the frequency of the auxiliary optical mode, and

$$
\Delta=\Omega_A-\Omega_T.
$$

The parameter $\gamma$ sets the magnetic coupling scale. In terms of
the phonon magnetic moment, it is defined by the Zeeman splitting of
the transverse circular modes. The dimensionless factor $\eta$ is

$$
\eta
=
\frac12
\left(
\sqrt{\frac{\Omega_A}{\Omega_T}}
+
\sqrt{\frac{\Omega_T}{\Omega_A}}
\right),
$$

and accounts for the unequal frequencies of the transverse and
auxiliary modes in the magnetic matrix element.

The diagonal terms $\pm a$ describe the opposite magnetic shifts of
the circular modes generated by $B_z$. The off-diagonal terms
$\pm v$ describe the coupling of the transverse modes to the
auxiliary mode generated by $B_x$.

There is no direct $L$--$R$ matrix element in
Eq.~(\ref{eq:D3}). Nevertheless, an effective coupling between the
two transverse modes is generated through the virtual process

$$
L\rightarrow A\rightarrow R.
$$

We derive this coupling explicitly below.

\section{Elimination of the auxiliary mode}

We write an eigenvector of Eq.~(\ref{eq:D3}) as

$$
|\psi\rangle
=
c_L|L\rangle
+
c_R|R\rangle
+
c_A|A\rangle.
$$

In vector notation,

$$
|\psi\rangle
=
\begin{pmatrix}
c_L\\
c_R\\
c_A
\end{pmatrix}.
$$

The eigenvalue equation

$$
D_{\mathrm{eff}}|\psi\rangle
=
E|\psi\rangle
$$

gives the three coupled equations

$$
(\Omega_T+a-E)c_L-vc_A=0,
$$

$$
(\Omega_T-a-E)c_R+vc_A=0,
$$

and

$$
-vc_L+vc_R+(\Omega_A-E)c_A=0.
$$

The third equation can be solved exactly for the auxiliary-mode
amplitude,

$$
c_A
=
\frac{v(c_L-c_R)}
{\Omega_A-E}.
$$

This equation shows that the auxiliary mode follows the difference
between the amplitudes of the two transverse modes.

Substituting this result into the first equation gives

$$
(\Omega_T+a-E)c_L
-
\frac{v^2}{\Omega_A-E}
(c_L-c_R)
=
0.
$$

Rearranging,

$$
\left(
\Omega_T+a
-
\frac{v^2}{\Omega_A-E}
-E
\right)c_L
+
\frac{v^2}{\Omega_A-E}c_R
=
0.
$$

Similarly, substitution into the second equation gives

$$
(\Omega_T-a-E)c_R
+
\frac{v^2}{\Omega_A-E}
(c_L-c_R)
=
0,
$$

or

$$
\frac{v^2}{\Omega_A-E}c_L
+
\left(
\Omega_T-a
-
\frac{v^2}{\Omega_A-E}
-E
\right)c_R
=
0.
$$

The original three-mode eigenvalue problem is therefore equivalent to

\begin{equation}
\label{eq:Deff-exact}
D_{LR}^{\mathrm{eff}}(E)
\begin{pmatrix}
c_L\\
c_R
\end{pmatrix}
=
E
\begin{pmatrix}
c_L\\
c_R
\end{pmatrix},
\end{equation}

with the energy-dependent effective transverse matrix

\begin{equation}
\label{eq:Deff-energy}
D_{LR}^{\mathrm{eff}}(E)
=
\begin{pmatrix}
\Omega_T+a-\dfrac{v^2}{\Omega_A-E}
&
\dfrac{v^2}{\Omega_A-E}
\\[6pt]
\dfrac{v^2}{\Omega_A-E}
&
\Omega_T-a-\dfrac{v^2}{\Omega_A-E}
\end{pmatrix}.
\end{equation}

Up to this point no approximation has been made in eliminating
$c_A$.

For the two modes evolving from the transverse doublet,

$$
E\simeq\Omega_T.
$$

Therefore,

$$
\Omega_A-E
=
\Delta-(E-\Omega_T).
$$

When the auxiliary mode is sufficiently separated from the transverse
doublet,

$$
|E-\Omega_T|\ll|\Delta|,
$$

we can expand

$$
\frac{1}{\Omega_A-E}
=
\frac{1}{\Delta}
\frac{1}
{1-(E-\Omega_T)/\Delta}
\simeq
\frac{1}{\Delta}
+
\frac{E-\Omega_T}{\Delta^2}
+\cdots.
$$

Keeping the leading term gives

\begin{equation}
\label{eq:Deff-LR}
D_{LR}^{\mathrm{eff}}
=
\begin{pmatrix}
\Omega_T+a-\dfrac{v^2}{\Delta}
&
\dfrac{v^2}{\Delta}
\\[6pt]
\dfrac{v^2}{\Delta}
&
\Omega_T-a-\dfrac{v^2}{\Delta}
\end{pmatrix}
+\cdots.
\end{equation}

Equivalently,

$$
D_{LR}^{\mathrm{eff}}
=
\left(
\Omega_T-\frac{v^2}{\Delta}
\right)I
+
\begin{pmatrix}
a&w\\
w&-a
\end{pmatrix}
+\cdots,
$$

where we have introduced

$$
w=\frac{v^2}{\Delta}.
$$

The term

$$
-\frac{v^2}{\Delta}I
$$

is a common frequency shift of the two transverse modes and therefore
does not affect their eigenvectors. The transverse eigenvectors are
determined by the reduced matrix

\begin{equation}
\label{eq:HLR}
H_{LR}
=
\begin{pmatrix}
a&w\\
w&-a
\end{pmatrix}
=
a\sigma_z+w\sigma_x.
\end{equation}

This form makes the physical content transparent. The $a\sigma_z$
term favors the original circular eigenstates $|L\rangle$ and
$|R\rangle$, while the effective coupling $w\sigma_x$ rotates the
eigenvectors within the transverse subspace.

\section{Eigenvalues and eigenvectors of the effective transverse problem}

We now diagonalize Eq.~(\ref{eq:HLR}) explicitly.

The eigenvalues $\lambda$ relative to the common shift satisfy

$$
\det
\begin{pmatrix}
a-\lambda&w\\
w&-a-\lambda
\end{pmatrix}
=0.
$$

Therefore,

$$
(a-\lambda)(-a-\lambda)-w^2=0,
$$

which gives

$$
\lambda^2=a^2+w^2.
$$

The two eigenvalues are consequently

$$
\lambda_{\pm}
=
\pm\sqrt{a^2+w^2}.
$$

Restoring the common shift, the two transverse eigenvalues are

\begin{equation}
\label{eq:Epm}
E_{\pm}
=
\Omega_T
-
\frac{v^2}{\Delta}
\pm
\sqrt{
a^2+\frac{v^4}{\Delta^2}
}.
\end{equation}

The corresponding splitting is

$$
E_+-E_-
=
2\sqrt{
a^2+\frac{v^4}{\Delta^2}
}.
$$

For weak transverse mixing,

$$
\left|
\frac{v^2}{a\Delta}
\right|
\ll1,
$$

this becomes

$$
E_+-E_-
\simeq
2|a|
\left[
1+
\frac12
\left(
\frac{v^2}{a\Delta}
\right)^2
+\cdots
\right].
$$

Thus, to leading order in the weak-field expansion, the splitting of
the transverse pair remains controlled by $a\propto B_z$.

We now determine the eigenvectors of the effective transverse problem

$$
H_{LR}
=
\begin{pmatrix}
a & w\\
w & -a
\end{pmatrix},
\qquad
w=\frac{v^2}{\Delta}.
$$

In the absence of the indirect coupling, $w=0$, this matrix is already
diagonal in the circular basis. Its eigenvectors are simply $|L\rangle$
and $|R\rangle$, with eigenvalues $+a$ and $-a$, respectively. A
nonzero $w$ mixes these two states. Since $H_{LR}$ is a real symmetric
$2\times2$ matrix, its two orthonormal eigenvectors can be obtained by
a real rotation of the original $(|L\rangle,|R\rangle)$ basis.

We therefore introduce a rotation angle $\beta$ and write

$$
|\widetilde L\rangle
=
\cos\beta\,|L\rangle
+
\sin\beta\,|R\rangle,
$$

and

$$
|\widetilde R\rangle
=
-\sin\beta\,|L\rangle
+
\cos\beta\,|R\rangle.
$$

The notation is chosen such that

$$
|\widetilde L\rangle\rightarrow|L\rangle,
\qquad
|\widetilde R\rangle\rightarrow|R\rangle
$$

as $w\rightarrow0$. Thus, $|\widetilde L\rangle$ and
$|\widetilde R\rangle$ denote the magnetic eigenstates that evolve
continuously from the original circular transverse modes.

In matrix form, define

$$
U(\beta)
=
\begin{pmatrix}
\cos\beta & \sin\beta\\
-\sin\beta & \cos\beta
\end{pmatrix}.
$$

The rotated eigenvectors can then be written as

$$
\begin{pmatrix}
|\widetilde L\rangle\\
|\widetilde R\rangle
\end{pmatrix}
=
U(\beta)
\begin{pmatrix}
|L\rangle\\
|R\rangle
\end{pmatrix}.
$$

Because $U(\beta)$ is orthogonal,

$$
U^{-1}(\beta)=U^T(\beta),
$$

so the effective matrix in the rotated basis is

$$
\widetilde H_{LR}
=
U(\beta)
H_{LR}
U^T(\beta).
$$

To determine $\beta$, we evaluate this product explicitly. First,

$$
U(\beta)H_{LR}
=
\begin{pmatrix}
\cos\beta & \sin\beta\\
-\sin\beta & \cos\beta
\end{pmatrix}
\begin{pmatrix}
a&w\\
w&-a
\end{pmatrix},
$$

which gives

$$
U(\beta)H_{LR}
=
\begin{pmatrix}
a\cos\beta+w\sin\beta
&
w\cos\beta-a\sin\beta
\\
w\cos\beta-a\sin\beta
&
-a\cos\beta-w\sin\beta
\end{pmatrix}.
$$

Multiplication by $U^T(\beta)$ can be carried out explicitly. We have

$$
\widetilde H_{LR}
=
U(\beta)H_{LR}U^T(\beta),
$$

with

$$
U(\beta)H_{LR}
=
\begin{pmatrix}
a\cos\beta+w\sin\beta
&
w\cos\beta-a\sin\beta
\\
w\cos\beta-a\sin\beta
&
-a\cos\beta-w\sin\beta
\end{pmatrix},
$$

and

$$
U^T(\beta)
=
\begin{pmatrix}
\cos\beta&-\sin\beta\\
\sin\beta&\cos\beta
\end{pmatrix}.
$$

Therefore,

$$
\widetilde H_{LR}
=
\begin{pmatrix}
a\cos\beta+w\sin\beta
&
w\cos\beta-a\sin\beta
\\
w\cos\beta-a\sin\beta
&
-a\cos\beta-w\sin\beta
\end{pmatrix}
\begin{pmatrix}
\cos\beta&-\sin\beta\\
\sin\beta&\cos\beta
\end{pmatrix}.
$$

The $(1,1)$ element is

$$
(\widetilde H_{LR})_{11}
=
(a\cos\beta+w\sin\beta)\cos\beta
+
(w\cos\beta-a\sin\beta)\sin\beta.
$$

Expanding,

$$
(\widetilde H_{LR})_{11}
=
a\cos^2\beta
-a\sin^2\beta
+
2w\sin\beta\cos\beta.
$$

Using

$$
\cos^2\beta-\sin^2\beta=\cos(2\beta),
$$

and

$$
2\sin\beta\cos\beta=\sin(2\beta),
$$

we obtain

$$
(\widetilde H_{LR})_{11}
=
a\cos(2\beta)+w\sin(2\beta).
$$

The $(1,2)$ element is

$$
(\widetilde H_{LR})_{12}
=
-(a\cos\beta+w\sin\beta)\sin\beta
+
(w\cos\beta-a\sin\beta)\cos\beta.
$$

Expanding,

$$
(\widetilde H_{LR})_{12}
=
w\cos^2\beta
-w\sin^2\beta
-
2a\sin\beta\cos\beta,
$$

and therefore

$$
(\widetilde H_{LR})_{12}
=
w\cos(2\beta)-a\sin(2\beta).
$$

Since the original matrix is symmetric and the transformation is
orthogonal,

$$
(\widetilde H_{LR})_{21}
=
(\widetilde H_{LR})_{12},
$$

so

$$
(\widetilde H_{LR})_{21}
=
w\cos(2\beta)-a\sin(2\beta).
$$

Finally, the $(2,2)$ element is

$$
(\widetilde H_{LR})_{22}
=
-(w\cos\beta-a\sin\beta)\sin\beta
+
(-a\cos\beta-w\sin\beta)\cos\beta.
$$

Expanding,

$$
(\widetilde H_{LR})_{22}
=
-a\cos^2\beta
+a\sin^2\beta
-
2w\sin\beta\cos\beta,
$$

which gives

$$
(\widetilde H_{LR})_{22}
=
-a\cos(2\beta)-w\sin(2\beta).
$$

Collecting the four matrix elements,

$$
\widetilde H_{LR}
=
\begin{pmatrix}
a\cos(2\beta)+w\sin(2\beta)
&
w\cos(2\beta)-a\sin(2\beta)
\\
w\cos(2\beta)-a\sin(2\beta)
&
-a\cos(2\beta)-w\sin(2\beta)
\end{pmatrix}.
$$

The rotated states are eigenvectors of $H_{LR}$ when this matrix is
diagonal. We therefore require its off-diagonal matrix element to
vanish,

$$
w\cos(2\beta)-a\sin(2\beta)=0.
$$

Rearranging gives

$$
w\cos(2\beta)=a\sin(2\beta),
$$

and, provided $a\neq0$,

$$
\tan(2\beta)=\frac{w}{a}.
$$

Since $w=v^2/\Delta$, this becomes

\begin{equation}
\label{eq:beta-condition}
\tan(2\beta)
=
\frac{v^2}{a\Delta}.
\end{equation}

It is convenient to define the dimensionless mixing parameter

\begin{equation}
\label{eq:t}
t
=
\frac{w}{a}
=
\frac{v^2}{a\Delta}.
\end{equation}

The rotation angle is therefore

\begin{equation}
\label{eq:beta-exact}
\beta
=
\frac{1}{2}\arctan t
=
\frac{1}{2}
\arctan\left(
\frac{v^2}{a\Delta}
\right).
\end{equation}

The physical meaning of $t$ can be seen directly from the two terms in
$H_{LR}$. The quantity $2a$ is the energy separation between the
unmixed $L$ and $R$ modes, whereas $w=v^2/\Delta$ is the indirect
coupling generated by the virtual process

$$
L\rightarrow A\rightarrow R.
$$

The ratio $w/a$ therefore measures the strength of this induced
coupling relative to the splitting that distinguishes the two circular
modes. When

$$
|w|\ll|a|,
$$

the two eigenvectors remain close to $|L\rangle$ and $|R\rangle$.
When $w$ becomes comparable to $a$, the rotation can no longer be
treated perturbatively.

Using Eq.~(\ref{eq:beta-condition}), the diagonal elements of the
rotated matrix can also be simplified. Since

$$
\cos(2\beta)
=
\frac{a}{\sqrt{a^2+w^2}},
\qquad
\sin(2\beta)
=
\frac{w}{\sqrt{a^2+w^2}},
$$

we obtain

$$
a\cos(2\beta)+w\sin(2\beta)
=
\sqrt{a^2+w^2}.
$$

Therefore,

$$
\widetilde H_{LR}
=
\begin{pmatrix}
\sqrt{a^2+w^2}&0\\
0&-\sqrt{a^2+w^2}
\end{pmatrix}.
$$

Restoring the common shift that was removed from
Eq.~(\ref{eq:Deff-LR}), the two transverse eigenvalues are consequently

\begin{equation}
\label{eq:Epmsm}
E_{\widetilde L,\widetilde R}
=
\Omega_T-\frac{v^2}{\Delta}
\pm
\sqrt{
a^2+\frac{v^4}{\Delta^2}
}.
\end{equation}

Thus the same rotation simultaneously gives the eigenvectors and the
modified splitting of the transverse doublet.

For weak transverse mixing,

$$
|t|
=
\left|
\frac{v^2}{a\Delta}
\right|
\ll1,
$$

the rotation angle can be expanded using

$$
\arctan t
=
t-\frac{t^3}{3}+\frac{t^5}{5}+\cdots.
$$

then 

\begin{equation}
\label{eq:beta-expansion}
\beta
=
\frac{t}{2}
-
\frac{t^3}{6}
+
\frac{t^5}{10}
+\cdots.
\end{equation}

To leading order,

$$
\beta
\simeq
\frac{t}{2}
=
\frac{v^2}{2a\Delta}.
$$

The rotated eigenvectors therefore become

$$
|\widetilde L\rangle
\simeq
|L\rangle
+
\frac{v^2}{2a\Delta}|R\rangle,
$$

and

$$
|\widetilde R\rangle
\simeq
|R\rangle
-
\frac{v^2}{2a\Delta}|L\rangle.
$$

This reproduces the $L$--$R$ mixing that would be obtained from
second-order nondegenerate perturbation theory. The effective
two-dimensional treatment, however, provides a more useful
interpretation: the perturbative coefficient

$$
\frac{v^2}{2a\Delta}
$$

is simply the small angle limit of the exact rotation of the transverse
eigenvectors.

It also makes the validity of this approximation explicit. The first
neglected term in Eq.~(\ref{eq:beta-expansion}) is proportional to
$t^3$, so its size relative to the leading term is of order $t^2$.
The expansion of the transverse eigenvectors is therefore controlled
by

$$
\left|
\frac{v^2}{a\Delta}
\right|
\ll1.
$$

\section{Full three-mode eigenvectors}

The rotation above determines the $L$--$R$ part of the eigenvectors.
The auxiliary-mode component can then be reconstructed from

$$
c_A
=
\frac{v(c_L-c_R)}
{\Omega_A-E}.
$$

To leading order in $v/\Delta$, we may replace the denominator by
$\Delta$.

For the state evolving from $|L\rangle$,

$$
c_L=\cos\beta,
\qquad
c_R=\sin\beta,
$$

and hence

$$
c_A^{(L)}
\simeq
\frac{v}{\Delta}
\left(
\cos\beta-\sin\beta
\right).
$$

Therefore,

\begin{equation}
\label{eq:L-full}
|\widetilde L\rangle
=
\cos\beta\,|L\rangle
+
\sin\beta\,|R\rangle
+
\frac{v}{\Delta}
\left(
\cos\beta-\sin\beta
\right)
|A\rangle
+\cdots.
\end{equation}

For the state evolving from $|R\rangle$,

$$
c_L=-\sin\beta,
\qquad
c_R=\cos\beta,
$$

so that

$$
c_A^{(R)}
\simeq
-\frac{v}{\Delta}
\left(
\cos\beta+\sin\beta
\right).
$$

Hence,

\begin{equation}
\label{eq:R-full}
|\widetilde R\rangle
=
-\sin\beta\,|L\rangle
+
\cos\beta\,|R\rangle
-
\frac{v}{\Delta}
\left(
\cos\beta+\sin\beta
\right)
|A\rangle
+\cdots.
\end{equation}

In the weak-mixing limit,

$$
\cos\beta\simeq1,
\qquad
\sin\beta\simeq\beta,
$$

and Eqs.~(\ref{eq:L-full}) and (\ref{eq:R-full}) reduce to

$$
|\widetilde L\rangle
\simeq
|L\rangle
+
\beta|R\rangle
+
\frac{v}{\Delta}(1-\beta)|A\rangle
+\cdots,
$$

and

$$
|\widetilde R\rangle
\simeq
|R\rangle
-
\beta|L\rangle
-
\frac{v}{\Delta}(1+\beta)|A\rangle
+\cdots.
$$

These expressions contain both physical mechanisms relevant to the
geometric matrix element: the direct admixture of the auxiliary mode
and the indirect rotation within the transverse $L$--$R$ subspace.

\section{Comparison with second-order perturbation theory}

The same transverse mixing can be obtained directly from standard
nondegenerate perturbation theory applied to the original $3\times3$
problem. This provides a useful check of the effective two-mode
description.

We write

$$
D_{\mathrm{eff}}=D_0+V,
$$

with

$$
D_0
=
\begin{pmatrix}
\Omega_T+a&0&0\\
0&\Omega_T-a&0\\
0&0&\Omega_A
\end{pmatrix},
\qquad
V
=
\begin{pmatrix}
0&0&-v\\
0&0&v\\
-v&v&0
\end{pmatrix}.
$$

The unperturbed eigenvalues are

$$
E_L^{(0)}=\Omega_T+a,
\qquad
E_R^{(0)}=\Omega_T-a,
\qquad
E_A^{(0)}=\Omega_A.
$$

There is no direct matrix element between the two transverse modes,

$$
\langle R|V|L\rangle=0,
$$

so their mixing vanishes at first order. At first order, $|L\rangle$
and $|R\rangle$ only acquire an auxiliary-mode component. For the
state evolving from $|L\rangle$,

$$
|L^{(1)}\rangle
=
\frac{
\langle A|V|L\rangle
}{
E_L^{(0)}-E_A^{(0)}
}|A\rangle
=
\frac{-v}{-(\Delta-a)}|A\rangle,
$$

and therefore

$$
|L^{(1)}\rangle
=
\frac{v}{\Delta-a}|A\rangle.
$$

Similarly,

$$
|R^{(1)}\rangle
=
\frac{
\langle A|V|R\rangle
}{
E_R^{(0)}-E_A^{(0)}
}|A\rangle
=
-\frac{v}{\Delta+a}|A\rangle.
$$

For $|a|\ll|\Delta|$, these reduce to

$$
|L^{(1)}\rangle
\simeq
\frac{v}{\Delta}|A\rangle,
\qquad
|R^{(1)}\rangle
\simeq
-\frac{v}{\Delta}|A\rangle,
$$

which are precisely the direct auxiliary-mode components obtained
above.

The mixing between $|L\rangle$ and $|R\rangle$ first appears at
second order. For a nondegenerate state $|n\rangle$, the component
along a different state $|m\rangle$ is

$$
\langle m|n^{(2)}\rangle
=
\sum_{l\neq n}
\frac{
\langle m|V|l\rangle
\langle l|V|n\rangle
}{
\left(E_n^{(0)}-E_m^{(0)}\right)
\left(E_n^{(0)}-E_l^{(0)}\right)
}.
$$

For the state evolving from $|L\rangle$, the only path that generates
an $|R\rangle$ component is

$$
L\rightarrow A\rightarrow R.
$$

Hence,

$$
\langle R|L^{(2)}\rangle
=
\frac{
\langle R|V|A\rangle
\langle A|V|L\rangle
}{
\left(E_L^{(0)}-E_R^{(0)}\right)
\left(E_L^{(0)}-E_A^{(0)}\right)
}.
$$

Using

$$
\langle R|V|A\rangle=v,
\qquad
\langle A|V|L\rangle=-v,
$$

together with

$$
E_L^{(0)}-E_R^{(0)}=2a,
\qquad
E_L^{(0)}-E_A^{(0)}=-(\Delta-a),
$$

gives

$$
\langle R|L^{(2)}\rangle
=
\frac{v^2}{2a(\Delta-a)}.
$$

For $|a|\ll|\Delta|$,

$$
\langle R|L^{(2)}\rangle
\simeq
\frac{v^2}{2a\Delta}.
$$

Similarly, the $|L\rangle$ component of the state evolving from
$|R\rangle$ is generated by

$$
R\rightarrow A\rightarrow L,
$$

and is

$$
\langle L|R^{(2)}\rangle
=
-\frac{v^2}{2a(\Delta+a)}
\simeq
-\frac{v^2}{2a\Delta}.
$$

Therefore, neglecting normalization corrections along the original
state, direct perturbation theory gives

\begin{equation}
\label{eq:PT-eigenvectors}
|\widetilde L\rangle_{\mathrm{PT}}
\simeq
|L\rangle
+
\frac{v^2}{2a\Delta}|R\rangle
+
\frac{v}{\Delta}|A\rangle,
\qquad
|\widetilde R\rangle_{\mathrm{PT}}
\simeq
|R\rangle
-
\frac{v^2}{2a\Delta}|L\rangle
-
\frac{v}{\Delta}|A\rangle.
\end{equation}

On the other hand, the effective $2\times2$ treatment gave

$$
\beta
=
\frac12
\arctan\left(\frac{v^2}{a\Delta}\right),
$$

which in the weak-mixing limit becomes

$$
\beta
\simeq
\frac{v^2}{2a\Delta}.
$$

Substituting this result into the eigenvectors derived above gives

$$
|\widetilde L\rangle
\simeq
|L\rangle
+
\frac{v^2}{2a\Delta}|R\rangle
+
\frac{v}{\Delta}
\left(
1-\frac{v^2}{2a\Delta}
\right)|A\rangle
+\cdots,
$$

and

$$
|\widetilde R\rangle
\simeq
|R\rangle
-
\frac{v^2}{2a\Delta}|L\rangle
-
\frac{v}{\Delta}
\left(
1+\frac{v^2}{2a\Delta}
\right)|A\rangle
+\cdots.
$$

In the weak transverse-mixing regime,

$$
|\beta|
\simeq
\left|
\frac{v^2}{2a\Delta}
\right|
\ll 1.
$$

The corrections proportional to $\beta$ in the auxiliary-mode
components are therefore small compared with their leading
contribution $v/\Delta$. Neglecting these relative
$\mathcal O(\beta)$ corrections gives

$$
|\widetilde L\rangle
\simeq
|L\rangle
+
\frac{v^2}{2a\Delta}|R\rangle
+
\frac{v}{\Delta}|A\rangle,
$$

and

$$
|\widetilde R\rangle
\simeq
|R\rangle
-
\frac{v^2}{2a\Delta}|L\rangle
-
\frac{v}{\Delta}|A\rangle,
$$

which agree directly with Eq.~(\ref{eq:PT-eigenvectors}).

Thus, the rotation of the transverse eigenvectors obtained by
eliminating the auxiliary mode is equivalent, in the weak-mixing
limit, to the second-order virtual process

$$
L\rightarrow A\rightarrow R.
$$

\section{Momentum dependence of the local phonon basis}

We next determine how the eigenvectors change when the propagation
direction is displaced from the reference direction.

For a small displacement along $k_x$, we write

$$
\theta\simeq\frac{k_x}{k_z}.
$$

The local longitudinal polarization is

$$
\mathbf e_A
=
(\sin\theta,0,\cos\theta),
$$

while the transverse spherical basis is

$$
\mathbf e_\theta
=
(\cos\theta,0,-\sin\theta),
\qquad
\mathbf e_\phi
=
(0,1,0).
$$

The circular transverse polarizations are

$$
\mathbf e_L
=
\frac{
\mathbf e_\theta+i\mathbf e_\phi
}{\sqrt2},
\qquad
\mathbf e_R
=
\frac{
\mathbf e_\theta-i\mathbf e_\phi
}{\sqrt2}.
$$

At $\theta=0$,

$$
\mathbf e_A=\hat{\mathbf z},
\qquad
\mathbf e_\theta=\hat{\mathbf x},
\qquad
\mathbf e_\phi=\hat{\mathbf y}.
$$

Since

$$
\partial_{k_x}
=
\frac{1}{k_z}\partial_\theta,
$$

we define

$$
G=\frac{1}{k_z}.
$$

At the reference direction,

$$
\partial_{k_x}|A\rangle
=
\frac{G}{\sqrt2}
\left(
|L\rangle+|R\rangle
\right),
$$

and

$$
\partial_{k_x}|L\rangle
=
-\frac{G}{\sqrt2}|A\rangle,
\qquad
\partial_{k_x}|R\rangle
=
-\frac{G}{\sqrt2}|A\rangle.
$$

Therefore,

\begin{equation}
\label{eq:basis-overlaps}
\braket{L|\partial_{k_x}A}
=
\frac{G}{\sqrt2},
\qquad
\braket{A|\partial_{k_x}R}
=
-\frac{G}{\sqrt2},
\qquad
\braket{L|\partial_{k_x}R}
=
0.
\end{equation}

\section{Momentum dependence of the magnetic couplings}

The magnetic field is fixed in the laboratory frame,

$$
\mathbf B
=
B_x\hat{\mathbf x}
+
B_z\hat{\mathbf z}.
$$

When the propagation direction rotates by $\theta$, its components in
the local phonon frame become

$$
B_\parallel
=
B_z\cos\theta+B_x\sin\theta,
$$

and

$$
B_\perp
=
B_x\cos\theta-B_z\sin\theta.
$$

Consequently,

$$
a(\theta)
=
\gamma
\left(
B_z\cos\theta+B_x\sin\theta
\right),
$$

and

$$
v(\theta)
=
\frac{\gamma\eta}{\sqrt2}
\left(
B_x\cos\theta-B_z\sin\theta
\right).
$$

At $\theta=0$,

$$
a=\gamma B_z,
\qquad
v=\frac{\gamma\eta}{\sqrt2}B_x.
$$

Their momentum derivatives are

\begin{equation}
\label{eq:mag-derivatives}
\partial_{k_x}a
=
G\gamma B_x,
\qquad
\partial_{k_x}v
=
-G\frac{\gamma\eta}{\sqrt2}B_z.
\end{equation}

\section{Derivative of the transverse rotation}

The field-induced transverse rotation contributes to the geometric
matrix element through the momentum dependence of $\beta$.

From Eq.~(\ref{eq:beta-exact}),

$$
\beta
=
\frac12\arctan t,
$$

so

\begin{equation}
\label{eq:dbeta-exact}
\partial_{k_x}\beta
=
\frac{
\partial_{k_x}t
}{
2(1+t^2)
}.
\end{equation}

Using

$$
t=\frac{v^2}{a\Delta},
$$

and neglecting the momentum dependence of $\Delta$ within the
three-mode model,

$$
\partial_{k_x}t
=
\frac{2v\,\partial_{k_x}v}{a\Delta}
-
\frac{v^2\,\partial_{k_x}a}{a^2\Delta}.
$$

Equivalently,

$$
\partial_{k_x}t
=
t
\left(
2\frac{\partial_{k_x}v}{v}
-
\frac{\partial_{k_x}a}{a}
\right).
$$

Using Eq.~(\ref{eq:mag-derivatives}),

$$
\frac{\partial_{k_x}v}{v}
=
-G\frac{B_z}{B_x},
$$

and

$$
\frac{\partial_{k_x}a}{a}
=
G\frac{B_x}{B_z}.
$$

Thus,

\begin{equation}
\label{eq:dt}
\partial_{k_x}t
=
-Gt
\left[
2\frac{B_z}{B_x}
+
\frac{B_x}{B_z}
\right].
\end{equation}

For $|t|\ll1$,

$$
\partial_{k_x}\beta
=
\frac12\partial_{k_x}t
+
\mathcal O(t^2\partial_{k_x}t).
$$

Using

$$
t
=
\frac{v^2}{a\Delta}
=
\frac{\gamma\eta^2}{2\Delta}
\frac{B_x^2}{B_z},
$$

we obtain

\begin{equation}
\label{eq:dbeta-expanded}
-\partial_{k_x}\beta
=
G\frac{\gamma\eta^2}{2\Delta}B_x
+
G\frac{\gamma\eta^2}{4\Delta}
\frac{B_x^3}{B_z^2}
+\cdots.
\end{equation}

This result shows explicitly why a process that originates from
second order in the coupling $v$ contributes already to the leading
field dependence of the current. The factor $1/a\propto1/B_z$ in the
transverse rotation compensates one power of magnetic field.

\section{Geometric matrix element}

The geometric matrix element is

$$
\mathcal A_x
=
\braket{
\widetilde L|
\partial_{k_x}
\widetilde R
}.
$$

There are two contributions at the order of interest.

The first originates from the direct auxiliary-mode components of
Eqs.~(\ref{eq:L-full}) and (\ref{eq:R-full}). To leading order in
$v/\Delta$ and $\beta$,

$$
\mathcal A_x^{(A)}
=
-\frac{v}{\Delta}
\braket{L|\partial_{k_x}A}
+
\frac{v}{\Delta}
\braket{A|\partial_{k_x}R}.
$$

Using Eq.~(\ref{eq:basis-overlaps}),

$$
\mathcal A_x^{(A)}
=
-\frac{\sqrt2Gv}{\Delta},
$$

and hence

\begin{equation}
\label{eq:A-direct}
\mathcal A_x^{(A)}
=
-G\frac{\gamma\eta}{\Delta}B_x.
\end{equation}

The second contribution arises from the momentum dependence of the
transverse rotation itself,

$$
\mathcal A_x^{(\beta)}
=
-\partial_{k_x}\beta.
$$

Using Eq.~(\ref{eq:dbeta-expanded}),

$$
\mathcal A_x^{(\beta)}
=
G\frac{\gamma\eta^2}{2\Delta}B_x
+
G\frac{\gamma\eta^2}{4\Delta}
\frac{B_x^3}{B_z^2}
+\cdots.
$$

Adding the two contributions gives

\begin{equation}
\label{eq:Ax-total}
\mathcal A_x
=
-G
\frac{\gamma\eta(2-\eta)}
{2\Delta}
B_x
+
G
\frac{\gamma\eta^2}
{4\Delta}
\frac{B_x^3}{B_z^2}
+\cdots.
\end{equation}

The first term contains contributions from both the direct
auxiliary-mode admixture and the induced transverse rotation.
Therefore, it should be regarded as the complete bilinear contribution
at this order, rather than as a ``first-order eigenvector''
contribution.

\section{Leading current and angular correction}

To leading order in the weak-field regime, the splitting of the two
transverse modes is proportional to $B_z$,

$$
\omega_2^2-\omega_1^2
\propto
B_z.
$$

Substitution of Eq.~(\ref{eq:Ax-total}) into
Eq.~(\ref{eq:J-eigenvector}) therefore gives a current of the form

\begin{equation}
\label{eq:J12}
J_x
=
J_1+J_2+\cdots,
\qquad
J_1=bB_xB_z,
\qquad
J_2=c\frac{B_x^3}{B_z}.
\end{equation}

Since both terms contain the same coherence factor and the same
leading transverse-mode splitting, their ratio is determined by the
relative coefficients in Eq.~(\ref{eq:Ax-total}). Therefore,

\begin{equation}
\label{eq:cb}
\frac{c}{b}
=
-\frac{\eta}{2(2-\eta)}.
\end{equation}

It follows that

\begin{equation}
\label{eq:J2J1}
\frac{J_2}{J_1}
=
-\frac{\eta}{2(2-\eta)}
\left(
\frac{B_x}{B_z}
\right)^2.
\end{equation}

\section{Comparison with the numerical dynamical-matrix calculation}
The parameters entering the analytical three-mode model are obtained
from first principles lattice dynamical calculations for SnTe. The
equilibrium structure and interatomic force constants were calculated
using density functional theory as implemented in VASP. The phonon
dynamical matrix was subsequently constructed using \textsc{Phonopy}
from finite-displacement calculations in a $3\times3\times3$
supercell. The resulting force constants were used to obtain the
phonon frequencies and polarization vectors and, in particular, to
identify the transverse doublet and the auxiliary optical mode entering
Eq.~(\ref{eq:D3}). The corresponding frequencies, $\Omega_T$ and
$\Omega_A$, determine the frequency-renormalization factor

$$
\eta
=
\frac{1}{2}
\left(
\sqrt{\frac{\Omega_A}{\Omega_T}}
+
\sqrt{\frac{\Omega_T}{\Omega_A}}
\right).
$$

For SnTe, the calculated phonon frequencies give

\begin{equation}
\label{eq:eta-SnTe}
\eta_{\mathrm{SnTe}}=1.04995858.
\end{equation}

The analytical three-mode model therefore predicts

$$
\left(\frac{c}{b}\right)_{\mathrm{theory}}
=
-\frac{\eta_{\mathrm{SnTe}}}
{2(2-\eta_{\mathrm{SnTe}})}
=
-0.55258569.
$$

The numerical calculation does not use the perturbative eigenvectors
derived above. Instead, the full field dependent dynamical matrix is
constructed and diagonalized numerically for each wave vector and each
magnetic field configuration. The field dependent eigenfrequencies and
polarization vectors obtained from this diagonalization are then
inserted directly into the dynamical matrix current expression,
Eq.~(\ref{eq:J-D}). In this way, the numerical current retains the
full magnetic field dependence of the phonon eigenvectors within the
dynamical matrix model, rather than assuming the weak-field expansion
used to derive $J_1$ and $J_2$. The comparison below therefore provides
a direct test of the regime in which the analytical three-mode
expansion reproduces the numerical dynamical-matrix result.
The analytical result can be tested against the current obtained
directly from Eq.~(\ref{eq:J-D}). We define

$$
q=\frac{B_x}{B_z}.
$$

At fixed $q$,

$$
B_z=\frac{B_x}{q},
$$

and Eq.~(\ref{eq:J12}) becomes

$$
J_x
=
\frac{b}{q}B_x^2
+
cqB_x^2.
$$

Thus, the current at each fixed field orientation has the form

\begin{equation}
\label{eq:Aq}
J_x=A_qB_x^2,
\end{equation}

where

$$
A_q
=
\frac{b}{q}+cq.
$$

\begin{figure}[h]
    \centering
    \includegraphics[width=1\textwidth]{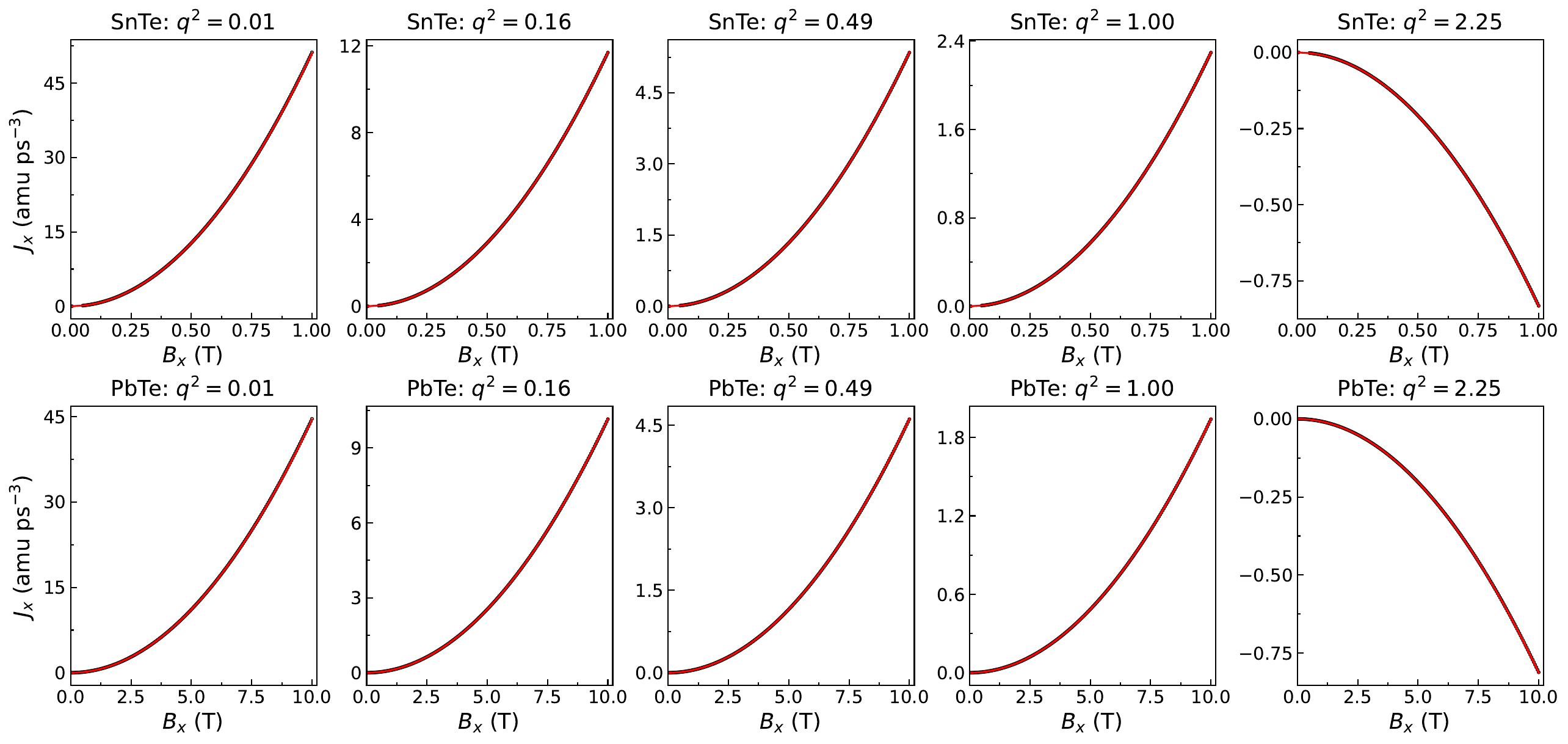}
    \caption{
Numerical transverse current obtained from the full magnetic
dynamical matrix for different fixed field ratios $q=B_x/B_z$.
The upper row shows results for SnTe with $g^*=0.5$, and the lower
row for PbTe with $g^*=0.043$. For each value of $q$, the magnetic field
magnitude is varied while the field orientation is held fixed.
The symbols show the numerical current obtained from
Eq.~(\ref{eq:J-D}), while the corresponding curves show fits to
$J_x=A_qB_x^2$. The quadratic dependence on $B_x$ is the expected
leading weak-field scaling at fixed $B_x/B_z$. The fitted
coefficients $A_q$ are subsequently used to determine the
relative weights of the $J_1$ and $J_2$ contributions.
    }
    \label{fig:S1}
\end{figure}

The first step of the numerical comparison is shown in
Fig.~\ref{fig:S1}. For each fixed value of the field ratio
$q=B_x/B_z$, the current is calculated directly from the full magnetic
dynamical matrix using Eq.~(\ref{eq:J-D}). The magnitude of the
magnetic field is varied while keeping $q$ fixed, and the resulting
current is fitted to Eq.~(\ref{eq:Aq}). The fitted coefficient $A_q$
therefore contains the angular dependence of the leading weak-field
current.

To separate the two angular contributions contained in $A_q$, we
multiply by $q$. Using

$$
A_q
=
\frac{b}{q}+cq,
$$

we obtain

\begin{equation}
\label{eq:qAq}
qA_q=b+cq^2.
\end{equation}

Thus, the intercept of $qA_q$ as a function of $q^2$ determines the
coefficient $b$ of the leading contribution

$$
J_1=bB_xB_z,
$$

whereas the slope determines the coefficient $c$ of

$$
J_2=c\frac{B_x^3}{B_z}.
$$

It is useful to remove the overall current scale $b$ and consider the
relative correction

\begin{equation}
\label{eq:Rnum}
R_{\mathrm{num}}(q^2)
=
\frac{qA_q}{b}-1
=
\frac{c}{b}q^2.
\end{equation}

The analytical three-mode calculation gives

\begin{equation}
\label{eq:Rtheory}
R_{\mathrm{theory}}(q^2)
=
-\frac{\eta}{2(2-\eta)}q^2.
\end{equation}

\begin{figure}[h]
    \centering
    \includegraphics[width=0.95\textwidth]{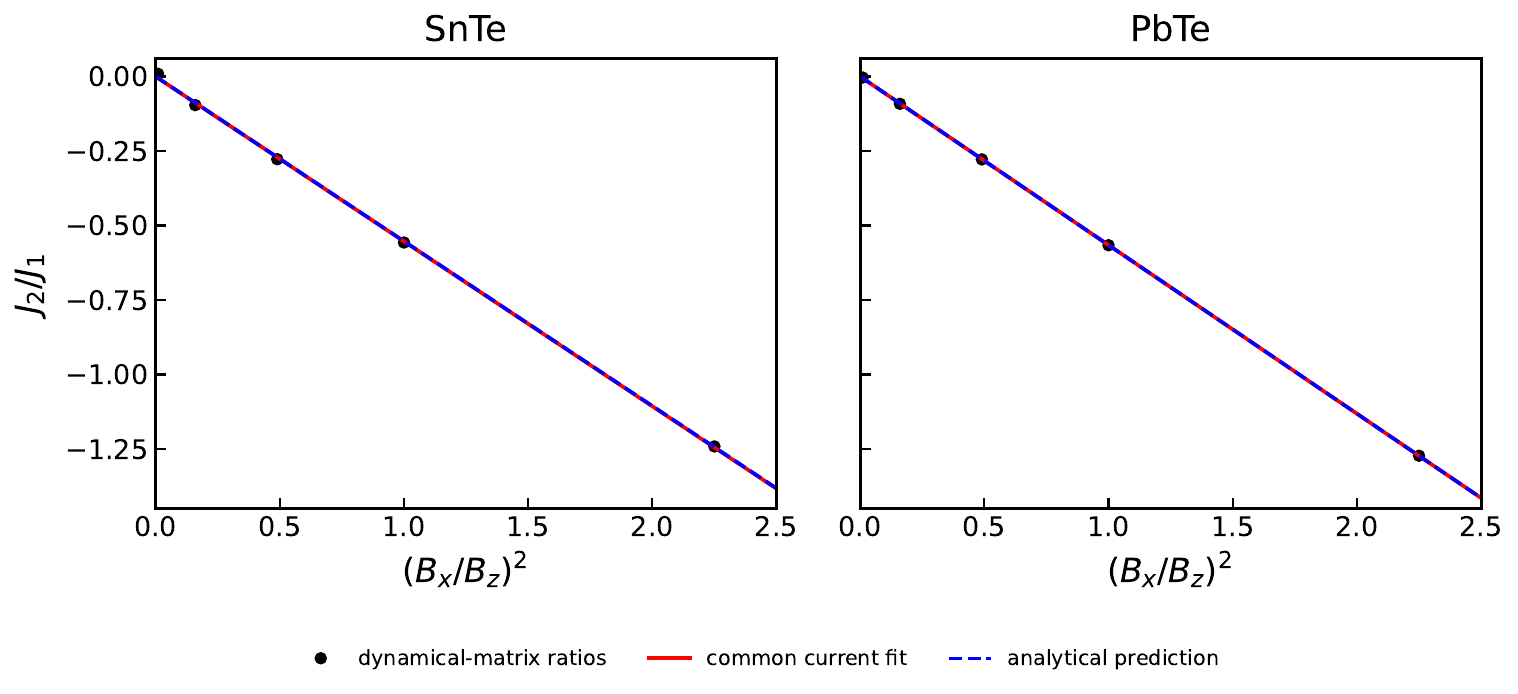}
    \caption{
    Relative angular correction to the transverse current. The symbols
    show
    $R_{\mathrm{num}}=qA_q/b-1$,
    obtained from the coefficients $A_q$ extracted from the
    dynamical-matrix calculations in Fig.~\ref{fig:S1}. The solid line
    shows a free linear fit,
    $R_{\mathrm{num}}=(c/b)_{\mathrm{fit}}q^2$, while the dashed line
    shows the analytical three-mode prediction,
    $R_{\mathrm{theory}}
    =-\eta[2(2-\eta)]^{-1}q^2$.
    Agreement between the two slopes tests the predicted relative
    coefficient of $J_2$ with respect to $J_1$.
    }
    \label{fig:S2}
\end{figure}

The second stage of the comparison is shown in
Fig.~\ref{fig:S2}. The numerical values extracted from the fits in
Fig.~\ref{fig:S1} are converted to the dimensionless quantity
$R_{\mathrm{num}}$. A free linear fit of these values determines the
numerical value of $c/b$, which can then be compared directly with the
analytical prediction in Eq.~(\ref{eq:Rtheory}).

This should be distinguished from a comparison of the absolute
coefficient $b$. Figure~\ref{fig:S2} tests the relative angular
dependence of the current, rather than its absolute amplitude.

The same result also provides a quantitative criterion for the range
over which the simpler bilinear form

$$
J_x\simeq J_1=bB_xB_z
$$

is sufficient. For SnTe,

$$
\left|
\frac{J_2}{J_1}
\right|
=
\frac{\eta}{2(2-\eta)}
\left(
\frac{B_x}{B_z}
\right)^2
=
0.55258569
\left(
\frac{B_x}{B_z}
\right)^2.
$$

Table~\ref{tab:BxBz-validity} shows the magnitude of the correction
for representative field ratios. The percentage in the third column
is the size of $J_2$ relative to $J_1$; it therefore directly measures
the error associated with retaining only the $B_xB_z$ contribution
within the present two-term expansion.

\begin{table}[h]
\centering
\begin{tabular}{c c c c}
\hline
$B_x/B_z$
&
$(B_x/B_z)^2$
&
$|J_2/J_1|$
&
Interpretation
\\
\hline
0.1 & 0.01 & 0.0055 \; (0.55\%) & $B_xB_z$ form essentially exact \\
0.2 & 0.04 & 0.0221 \; (2.21\%) & very good approximation \\
0.3 & 0.09 & 0.0498 \; (4.98\%) & approximately 5\% correction \\
0.4 & 0.16 & 0.0885 \; (8.85\%) & approximately 10\% correction \\
0.5 & 0.25 & 0.1383 \; (13.83\%) & noticeable correction \\
0.7 & 0.49 & 0.2711 \; (27.11\%) & $J_2$ is significant \\
1.0 & 1.00 & 0.5532 \; (55.32\%) & $B_xB_z$ term alone insufficient \\
1.5 & 2.25 & 1.2447 \; (124.47\%) & $J_2$ exceeds $J_1$ \\
\hline
\end{tabular}
\caption{
Magnitude of the first angular correction to the bilinear
$B_xB_z$ current for SnTe, using
$\eta=1.04995858$. The quantity $|J_2/J_1|$ measures the relative
importance of the $B_x^3/B_z$ contribution compared with the
$B_xB_z$ term. The ratios are calculated from the regenerated common fit
$c/b=-0.55318773$ for $g=0.5$ and $B_x\leq 3\,\mathrm{T}$.
The table characterizes the validity of retaining $J_1$ alone and
should not be interpreted as the validity range of the full
three-mode weak-field expansion.
}
\label{tab:BxBz-validity}
\end{table}

For example, requiring the correction to the $B_xB_z$ form to remain
below $5\%$ gives

$$
0.55258569
\left(
\frac{B_x}{B_z}
\right)^2
<0.05,
$$

or approximately

$$
\left|
\frac{B_x}{B_z}
\right|
<0.30.
$$

Similarly, a $10\%$ criterion gives

$$
\left|
\frac{B_x}{B_z}
\right|
<0.43.
$$

It is important that these bounds characterize only the range of
validity of the simple $J_x\propto B_xB_z$ approximation. They do not
imply that the three-mode description itself breaks down when
$J_2/J_1$ becomes large. For example, at $B_x/B_z=1$, the correction
is already approximately $55\%$, so $J_1$ alone is not a good
description, while the combined expression

$$
J_x
=
bB_xB_z
+
c\frac{B_x^3}{B_z}
$$

may still accurately reproduce the numerical dynamical-matrix result.
The validity of this more complete expression is instead controlled
by the small parameters associated with eliminating the auxiliary mode
and expanding the transverse rotation, namely $|a/\Delta|$,
$|v/\Delta|$, and $|v^2/(a\Delta)|$.

\section{Diagonal contribution and total phonon current}

We now add the diagonal contribution to the interband current derived
above. For a phonon eigenmode $n$, the diagonal energy current is
given by its energy multiplied by its group velocity,
\begin{equation}
    J_i^{\mathrm{diag}}
    =
    \frac{\hbar}{V_c}
    \sum_n
    \omega_n
    \partial_{k_i}\omega_n\,
    n_n .
    \label{eq:Jdiag_general}
\end{equation}
We consider the two circular transverse modes at
$\mathbf{k}=k\hat{\mathbf z}$. For a small rotation of the propagation
direction in the $xz$ plane,
\[
    \vartheta \simeq \frac{k_x}{k},
\]
the component of the magnetic field parallel to the propagation
direction is
\[
    B_\parallel
    =
    B_z\cos\vartheta+B_x\sin\vartheta.
\]
Consequently, at the reference direction,
\[
    \left.
    \partial_{k_x}B_\parallel
    \right|_{k_x=0}
    =
    \frac{B_x}{k}.
\]

The two circular modes carry opposite phonon magnetic moments. We
therefore write their frequencies, to first order in the magnetic
field, as
\[
    \Omega_{L/R}
    =
    \Omega_T
    \pm
    \frac{\mu_{\mathrm{ph}}}{2\hbar}B_\parallel .
\]
At $\mathbf{k}=k\hat{\mathbf z}$ this gives
\[
    \Omega_{L/R}
    =
    \Omega_T
    \pm
    \frac{\mu_{\mathrm{ph}}B_z}{2\hbar},
\]
while their transverse group velocities are
\[
    \partial_{k_x}\Omega_{L/R}
    =
    \pm
    \frac{\mu_{\mathrm{ph}}B_x}{2\hbar k}.
\]

A linearly polarized phonon contains equal occupations of the two
circular components,
\[
    n_L=n_R=\frac{1}{2}.
\]
Substitution into Eq.~\eqref{eq:Jdiag_general} gives
\[
    J_x^{\mathrm{diag}}
    =
    \frac{\hbar}{2V_c}
    \left(
        \Omega_L\partial_{k_x}\Omega_L
        +
        \Omega_R\partial_{k_x}\Omega_R
    \right).
\]
Using the expressions above,
\[
\begin{aligned}
    \Omega_L\partial_{k_x}\Omega_L
    &=
    \left(
        \Omega_T+
        \frac{\mu_{\mathrm{ph}}B_z}{2\hbar}
    \right)
    \frac{\mu_{\mathrm{ph}}B_x}{2\hbar k},
    \\
    \Omega_R\partial_{k_x}\Omega_R
    &=
    -
    \left(
        \Omega_T-
        \frac{\mu_{\mathrm{ph}}B_z}{2\hbar}
    \right)
    \frac{\mu_{\mathrm{ph}}B_x}{2\hbar k}.
\end{aligned}
\]
The terms proportional to $\Omega_TB_x$ cancel between the two
circular modes, leaving
\begin{equation}
    J_x^{\mathrm{diag}}
    =
    \frac{\mu_{\mathrm{ph}}^2B_xB_z}
         {4V_c\hbar k}.
    \label{eq:Jdiag_ph}
\end{equation}
The corresponding diagonal contribution along $y$ vanishes for the
chosen magnetic-field geometry,
\[
    J_y^{\mathrm{diag}}=0.
\]

We next combine this result with the interband contribution obtained
above. For a linearly polarized excitation at angle $\theta$,
\[
    |\psi_\theta\rangle
    =
    \frac{1}{\sqrt{2}}
    \left(
        e^{-i\theta}|L\rangle
        +
        e^{i\theta}|R\rangle
    \right),
\]
so that
\[
    \langle a_L^\dagger a_R\rangle
    =
    \frac{1}{2}e^{2i\theta}.
\]
Using the geometric matrix element derived in Eq.~(S24), the leading
interband current is
\begin{equation}
    J_x^{\mathrm{inter}}
    =
    \frac{\mu_{\mathrm{ph}}^2B_xB_z}
         {2V_c\hbar k}
    \eta(2-\eta)
    \frac{\Omega_T}{\Delta}
    \cos(2\theta),
    \qquad
    J_y^{\mathrm{inter}}
    =
    \frac{\mu_{\mathrm{ph}}^2B_xB_z}
         {2V_c\hbar k}
    \eta(2-\eta)
    \frac{\Omega_T}{\Delta}
    \sin(2\theta).
    \label{eq:Jinter_ph}
\end{equation}

Adding Eqs.~\eqref{eq:Jdiag_ph} and \eqref{eq:Jinter_ph}, the total
bare-phonon transverse energy current is
\[
    J_x^{E}
    =
    \frac{\mu_{\mathrm{ph}}^2B_xB_z}
         {2V_c\hbar k}
    \left[
        \frac{1}{2}
        +
        \eta(2-\eta)
        \frac{\Omega_T}{\Delta}
        \cos(2\theta)
    \right],
\]
and
\[
    J_y^{E}
    =
    \frac{\mu_{\mathrm{ph}}^2B_xB_z}
         {2V_c\hbar k}
    \eta(2-\eta)
    \frac{\Omega_T}{\Delta}
    \sin(2\theta).
\]

\section{Renormalization by phonon--photon hybridization}

We now extend the phonon current to a phonon-polariton branch with
frequency $\omega_k$. We regard the polariton frequency as a function
of the bare transverse phonon frequency,
\[
    \omega_k=\omega_k(\Omega_T),
\]
at fixed wavevector $k$. A small magnetic shift
$\delta\Omega_T$ of the bare phonon therefore produces the polariton
frequency shift
\[
    \delta\omega_k
    =
    \left.
    \frac{\partial\omega_k}{\partial\Omega_T}
    \right|_k
    \delta\Omega_T .
\]
We define the corresponding Zeeman-renormalization factor as
\begin{equation}
    h_k
    \equiv
    \left.
    \frac{\partial\omega_k}{\partial\Omega_T}
    \right|_k
    =
    \frac{\delta\omega_k^{\mathrm{pol}}}
         {\delta\Omega_T^{\mathrm{ph}}}.
    \label{eq:hk}
\end{equation}

For the two circular channels, the field-dependent polariton
frequencies are consequently
\[
    \omega_{k,L/R}
    =
    \omega_k
    \pm
    h_k
    \frac{\mu_{\mathrm{ph}}}{2\hbar}
    B_\parallel .
\]
At the reference direction this becomes
\[
    \omega_{k,L/R}
    =
    \omega_k
    \pm
    h_k
    \frac{\mu_{\mathrm{ph}}B_z}{2\hbar}.
\]

The transverse group velocity follows by differentiating the same
field-dependent polariton dispersion. To leading order in the
magnetic field,
\[
    \partial_{k_x}\omega_{k,L/R}
    =
    \pm
    h_k
    \frac{\mu_{\mathrm{ph}}}{2\hbar}
    \partial_{k_x}B_\parallel
    =
    \pm
    h_k
    \frac{\mu_{\mathrm{ph}}B_x}{2\hbar k},
\]
where derivatives of $h_k$ multiply higher-order corrections and are
neglected in the present leading-field expansion.

The diagonal current of an equally populated pair of circular
polariton modes is therefore
\[
    J_{x,k}^{\mathrm{diag}}
    =
    \frac{\hbar F_k}{2V_c}
    \left(
        \omega_{k,L}
        \partial_{k_x}\omega_{k,L}
        +
        \omega_{k,R}
        \partial_{k_x}\omega_{k,R}
    \right),
\]
where $F_k$ selects the phonon contribution to the energy current.
Substitution of the field-dependent frequencies gives
\[
\begin{aligned}
    J_{x,k}^{\mathrm{diag}}
    &=
    \frac{\hbar F_k}{2V_c}
    \Bigg[
    \left(
        \omega_k+
        h_k\frac{\mu_{\mathrm{ph}}B_z}{2\hbar}
    \right)
    h_k\frac{\mu_{\mathrm{ph}}B_x}{2\hbar k}
    \\
    &\hspace{3.0cm}
    -
    \left(
        \omega_k-
        h_k\frac{\mu_{\mathrm{ph}}B_z}{2\hbar}
    \right)
    h_k\frac{\mu_{\mathrm{ph}}B_x}{2\hbar k}
    \Bigg].
\end{aligned}
\]
The terms proportional to $\omega_kB_x$ again cancel. The remaining
term is
\begin{equation}
    J_{x,k}^{\mathrm{diag}}
    =
    \frac{F_k\mu_{\mathrm{ph}}^2B_xB_z}
         {4V_c\hbar k}
    h_k^2.
    \label{eq:Jdiag_pol}
\end{equation}
Thus the two factors of $h_k$ have a direct origin: one renormalizes
the Zeeman splitting of the polariton energy and the second
renormalizes the corresponding transverse group velocity.

We next consider the interband contribution. The bare-phonon result
in Eq.~\eqref{eq:Jinter_ph} contains one factor originating from the
splitting of the two circular modes and one factor originating from
the geometric matrix element. Under phonon--photon hybridization, the
former is renormalized according to Eq.~\eqref{eq:hk},
\[
    \delta\Omega_T^{\mathrm{ph}}
    \longrightarrow
    \delta\omega_k^{\mathrm{pol}}
    =
    h_k\delta\Omega_T^{\mathrm{ph}}.
\]
The energy carried by the excitation is likewise the polariton energy
$\hbar\omega_k$ rather than the bare phonon energy
$\hbar\Omega_T$. The phonon contribution is finally weighted by
$F_k$. Applying these replacements to Eq.~\eqref{eq:Jinter_ph} gives
\begin{equation}
    J_{x,k}^{\mathrm{inter}}
    =
    \frac{F_k\mu_{\mathrm{ph}}^2B_xB_z}
         {2V_c\hbar k}
    h_k
    \eta(2-\eta)
    \frac{\omega_k}{\Delta}
    \cos(2\theta),
    \qquad
    J_{y,k}^{\mathrm{inter}}
    =
    \frac{F_k\mu_{\mathrm{ph}}^2B_xB_z}
         {2V_c\hbar k}
    h_k
    \eta(2-\eta)
    \frac{\omega_k}{\Delta}
    \sin(2\theta).
    \label{eq:Jinter_pol}
\end{equation}

Combining Eqs.~\eqref{eq:Jdiag_pol} and
\eqref{eq:Jinter_pol}, and defining
\begin{equation}
    K_k
    =
    \frac{F_k\mu_{\mathrm{ph}}^2B_xB_z h_k}
         {2V_c\hbar k},
    \qquad
    C_k
    =
    \eta(2-\eta)\frac{\omega_k}{\Delta},
    \label{eq:KC}
\end{equation}
we obtain
\[
    \boxed{
    J_{x,k}^{E}
    =
    K_k
    \left[
        \frac{h_k}{2}
        +
        C_k\cos(2\theta)
    \right],
    \qquad
    J_{y,k}^{E}
    =
    K_k C_k\sin(2\theta)
    }.
\]
The longitudinal phonon contribution to the energy current follows
from the phonon fraction of the polariton energy and its group
velocity,
\[
    \boxed{
    J_{z,k}^{E}
    =
    \frac{\hbar}{V_c}
    F_k\omega_k v_k,
    \qquad
    v_k=\frac{d\omega_k}{dk}.
    }
\]
These expressions give the phonon contribution to the total
phonon-polariton energy current used in the main text.

\end{document}